\documentclass{article}
\usepackage{chemfig}
\usepackage{mathtools}
\usepackage[version=4]{mhchem}
\usepackage{chemfig}
\usepackage{amsfonts}
\usepackage{amsmath}
\usepackage{amssymb}
\usepackage{graphicx}
\usepackage{booktabs}
\providecommand{\U}[1]{\protect\rule{.1in}{.1in}}

\date{}
\begin{document}

\title{A Stochastic Compartmental Model for COVID-19}
\author{Giorgio SONNINO${}^1$, Fernando MORA${}^2$, and Pasquale NARDONE${}^1$\\
${}^1$Universit{\'e} Libre de Bruxelles (ULB), Facult{\'e} de Sciences\\
Bvd du Triomphe, Campus de la Plaine, CP 231, 1050 Brussels, Belgium\\
${}^2$Centro de Física No Lineal y Sistemas Complejos de Santiago (CFNL)\\
Vitacura 2902, Las Condes, Santiago, Chile\\
Emails: gsonnino@ulb.ac.be, fmora@unice.fr, pnardon@ulb.ac.be}
\maketitle
\begin{abstract}
\noindent We propose two stochastic models for the  Coronavirus pandemic. The statistical properties of the models, in particular the correlation functions and the probability density function, have duly been computed. Our models, which generalises a model previously proposed and published in a specialised journal, take into account the adoption of the lockdown measures as well as the crucial role of the hospitals and Health Care Institutes. To accomplish this work we have analysed two scenarios: the $SIS$-model (Susceptible $\Rightarrow$ Infectious $\Rightarrow$ Susceptible) in presence of the lockdown measures and the $SIS$-model integrated with the action of the hospitals (always in presence of the lockdown measures). We show that in the case of the pure $SIS$-model, once the lockdown measures are removed, the Coronavirus will start growing again. However, in the second scenario, beyond a certain threshold of the hospital capacities, the Coronavirus is not only kept under control, but its capacity to spread tends to diminish in time. Therefore, the combined effect of the lockdown measures with the action of the hospitals and health Institutes is able to contain and dampen the spread of the SARS-CoV-2 epidemic. This result can be used during a period of time when the massive distribution of delivery of a limited number of vaccines in a given population is not yet feasible. By way of example, we analysed the data for USA and France where the intensities of the noise have been estimated by Statistical Mechanics. In particular, for USA we have analysed two possible hypotheses: USA is still subject to the first wave of infection by  and USA is in the second (or third) wave of SARS-CoV-2 infection.The agreement between theoretical predictions and real data confirms the validity of our approach.
\vskip0.2cm
\noindent {\bf Key words}: Mathematical model; COVID-19; Dynamics of population; Pneumonia.
\end{abstract}

\section{\bf Introduction}\label{I}

\noindent The current number of infectious people by SARS-CoV-2 brought us back to the lockdown phase, but the tracking and swabs give a clearer vision of the situation. By reading the figures on new cases of positivity, it seems that we are returning back to mid-April 2020. Hospitalisations are also growing up including new hospitalisations in intensive care. In this respect, the situation appears deeply different compared to the first months of the epidemic, when the prolonged lockdown was necessary, which allowed to reduce the spread of the Coronavirus significantly. The percentage ratio between new cases and swabs has now stood at around 5$\%$ \cite{swword}. I is therefore far from almost 30 per cent of the "darkest" periods of the pandemic in certain European Countries. However, these data induce to increase further preventive strategies, focusing on tracing and beyond, taking into account that in some countries the "weight" of the pandemic is starting to be felt again on hospitals. So, we are in the middle of the second wave, but the virus circulates among humans and therefore the lockdown period has limited its spread, so much so that we have assisted to a sort of "stagnation" in the months of June and July for the subsequent resumption with August. However, \textit{the virus did not changed or that the clinical manifestations that it induces have not changed} \cite{cauda}. In this second period of COVID-19, we may trust much more of statistical data thanks to the increased number of swabs done and the tracing and monitoring of people who have contracted the SARS-CoV-2 infection. Of course, there are still some cases, especially of asymptomatic subjects who unfortunately can transmit the infection: about 20$\%$ of cases of contagion would in fact be referred to subjects who do not show any health problems. This situation also explains why the percentage of positive swabs in the first weeks of the lockdown was around 30 percent while today it is much lower. Anyway, we are still far from the figures, of several thousand, of last March-April, but the trend is still on the rise. For this reason, it is essential to contain the infections, also through "ad hoc" measures (lockdown, quarantine, isolation). The carelessness during the summer leave has created a clear epidemiological link, given that the cases resumed in August and now we have to remedy, also because there are signs coming from nearby France that warn under this aspect.

\noindent $\bullet$ \textit{USA situation}. After sustained declines in the number of COVID-19 cases over recent months, restrictions are starting to ease across the United States. Numbers of new cases are falling or stable at low numbers in some states, but they are surging in many others. Overall, the USA. is experiencing a sharp increase in the number of new cases a day, and by late June, had surpassed the peak rate of spread in early April. In the USA, cases are rising quickly right now. With more than 11 million confirmed cases, USA has the highest number of infections in the world and the spread of the Coronavirus shows no sign of slowing down. During the spring wave, swabs were mostly limited to confirming cases in people who were already in hospital, meaning the true scale of that outbreak wasn't fully captured. But the latest data compiled by the COVID Tracking Project shows the current surge is not just down to increased testing, the number of tests carried out in USA was up by 12.5$\%$ week on week, while the number of cases increased by more than 40$\%$ \cite{usasituation}. One likely cause is the change of season and colder weather driving people indoors to socialise, where the risk of spread is heightened due to less social distancing and poor ventilation. The strain on hospitals is growing. Because of the change in the level of testing, a better way to compare waves is to look at the number of people being admitted to hospital because of Covid-19. This data shows that roughly the same number of people across the USA were in hospital during the first and second waves of the outbreak. But there are already more people in hospital during the current wave - more than 70000 at the moment. New deaths per day have not begun to climb, but some hospitals’ intensive care units have recently reached full capacity \cite{usasituation1}. This wave is hitting every USA region. We can say that the USA as a whole is not in a second wave because the first wave never really stopped. The virus is simply spreading into new populations or resurgent in places that let down their guard too soon. This time around, every region is seeing a spike in new cases \cite{usasituation}.

\noindent $\bullet$ \textit{France situation}. Coronavirus cases in France rose dramatically, but the number of daily deaths declined. The total number of infections stands at around of 1829659, with the death toll standing at 42207, according to the Ministry of Health. The overall downward trend is a positive sign after Coronavirus numbers increased drastically by reaching a peak of 86852 cases. Some 99 departments throughout France remain in a vulnerable position with 3042 clusters of infection considered serious \cite{francesituation}. France now has the fourth highest number of infections in the world behind the USA, India and Brazil. Despite the French government’s efforts to slow down the spread of the virus with targeted measures such as lockdown and isolation, the rate of infections has continued to rise across the Country. Talk of a second wave, and how to avoid it, has been around since the end of the spring lockdown, but the current situation seems to suggest it was inevitable. One month after instituting the lockdown measures were applied in major cities starting from Paris and three weeks after extending it to almost the whole Country, France sees the light at the end of the dark tunnel of the second wave of Coronavirus. French Health Minister announced that he believes that the peak of the pandemic has passed, and the control has been regained. Indeed, looking at the data, after the peak of 86000 cases recorded on 7 November, the Country has moved on figures of around 30000 daily infections with a minimum level of 230000 recorded on 13 November. Figures that are returning in line with those of a month ago when the lockdown began \cite{francesituation1}. However, the Country will remain in a second nationwide lockdown that will continue until Dec. 1 unless the number of infections and deaths decrease significantly.

\noindent $\bullet$ \textit{Modelling the spread of SARS-CoV-2 infection}. Until vaccine arrives, the tried-and-true public health measures of the last months (social distancing, universal mask wearing, frequent hand-washing and avoiding crowded indoor spaces) are the ways to stop the first wave and thwart a second one. And when there are surges like what is happening now in the word, further reopening plans need to be put on hold. Meanwhile, in order to understand the situation, theoretic studies, mainly based on mathematical model with the use of up-to-date data, must be carried in order to understand the spread of this pandemic in future and provide predictions. In this study we propose a simple stochastic model with the objective to show how public health interventions can influence the outcome of the epidemic, by limiting and even damping the spread of the SARS-CoV-2 epidemic, waiting for the production and delivering of an effective vaccine.  The analysis is carried by assigning  population in different \textit{compartments} with labels, such as $S$, $I$, $R$ and $D$ (Susceptible, Infectious, Recovered, and Deaths). People can progress between the compartments. The order of the labels shows the flow patterns between the compartments; for example $SIS$ means susceptible, infectious and, therefore, Susceptible again. The origin of these models is at the beginning of the twentieth century~\cite{may}. The goal of our model is to predict how a disease spreads, or the dynamics of the number of infectious people, and the duration of the COVID-19 pandemic. The dynamics of this kind of models is governed by ordinary differential equations (O.D.E.s). Generally, these O.D.E.s are deterministic, but in this work the dynamics is described by a stochastic differential equation, which is more realistic. 

\noindent The work is organised as follows. In Section~(\ref{DM}), we study the deterministic $SIS$-model in presence of the lockdown measures. We shall see that as soon as the lockdown measures are stopped, the spread of the Coronavirus begins to grow back vigorously once a certain threshold is exceeded. In the Subsection~(1.2) we ask ourselves the following question: \textit{Let us suppose that a government needs a number $x$ of weeks (e.g. 6 weeks) before being able to deliver an effective vaccine. What is the minimum value of infectious people that should be expected, after the adoption of severe lockdown measures imposed on the population, to be sure that after $x$-weeks (e.g. after 6 weeks) the total number of the infectious individuals by Coronavirus remains below a threshold pre-set in advance, and acceptable from a medical and political point of view ?}. Subsection~(\ref{SW}) answers the above-mentioned question for the $SIS$-model subject to the lockdown measures. In Section~(\ref{SM}) we introduce, and study, the Stochastic $SIS$-model subject to the lockdown measures (the $(SIS)_L$-model). We compute the relevant correlation functions as well as the probability density function. Hospitals and health institutions play a crucial role in hindering the spread of the Coronavirus. In Section~(\ref{DM2}), we propose a model that accounts for people who are only traced back to hospitalised infectious individuals. This model is referred to as the $(SISI_hr_hd_h)_L$-model. We shall see that beyond a certain threshold of the hospital capacities, the coronavirus not only is kept under control, but its spread even tends to diminish in time. Therefore, the combined effect of the lockdown measures with efficient Hospitals and health Institutes is able to contain and dampen the spread of the SARS-CoV-2 epidemic, waiting for the delivery of a limited number of vaccines in a given population. In Section~(\ref{SMS}), we study the simplified $(SISI_h)_L$ stochastic model by computing the relevant correlation functions. Comparison between the theoretical predictions and the real data for USA and France can be found in Section~(\ref{comp}). Concluding remarks are reported in Section~(\ref{C}). An estimation of the intensity of the noise by Statistical Mechanics can be found in Annex 1. The derivation of the exact solution of Richards equation is reported in Appendix 2. Tables of real data for infectious individuals in USA and France are shown in Annex 3.

\section{\bf The Deterministic Model in Presence of the Lockdown Measures - The $(SIS)_L$-Model}\label{DM}
\begin{figure}[hbt!]
\hskip 2truecm
\includegraphics[width=8cm, height=6cm]{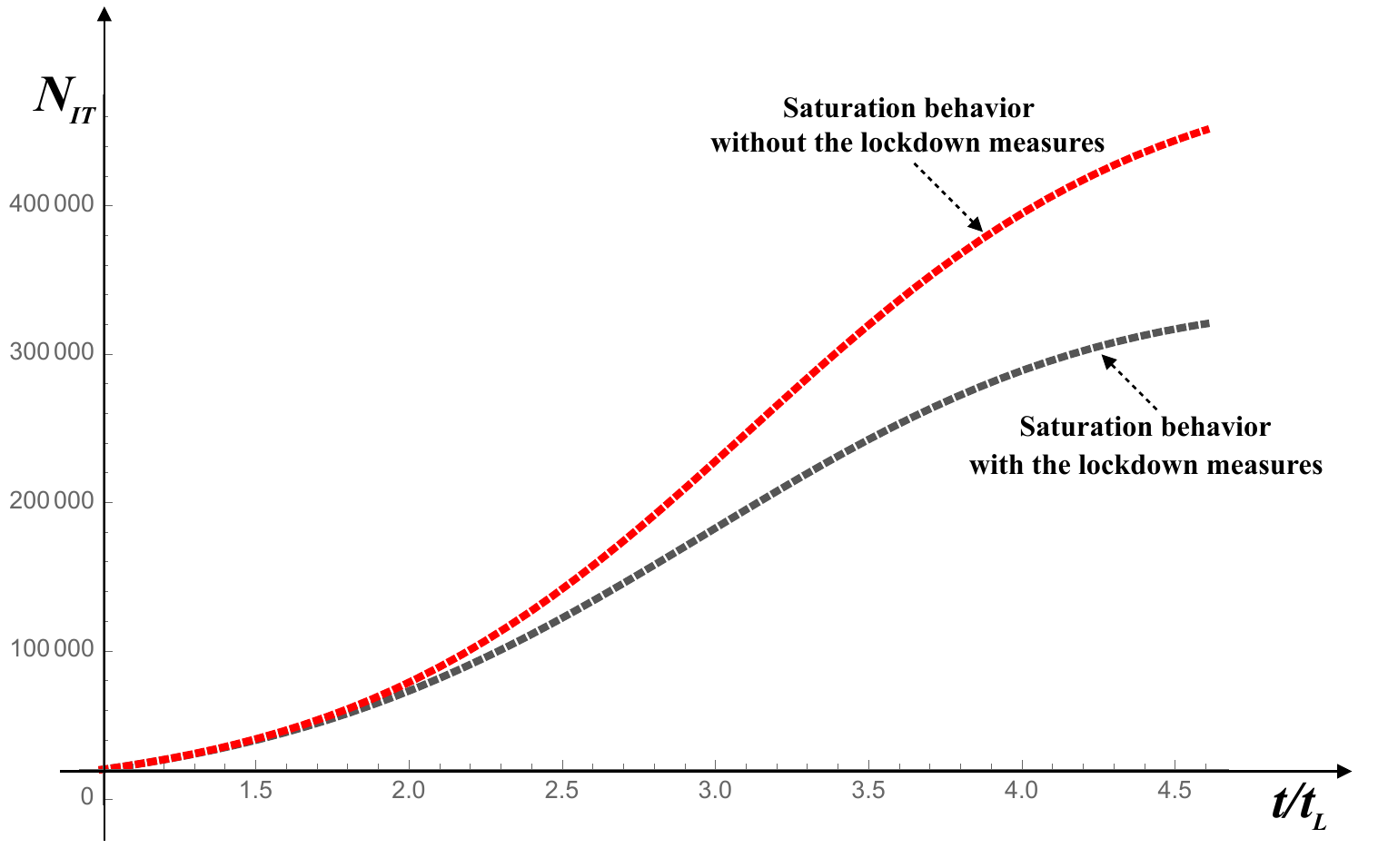}
\caption{
\textit{{\bf Solution of Eq.~(\ref{D11}) for Italy, first wave of infection by SARS-CoV-2.} The black and the red dotted lines refer to the solutions of Eq.~(\ref{D11}) with and without the application of the lockdown measures, respectively. The values of the parameters are $I_0$=5000, $K=150000$, $\alpha=1.5$, and $\beta=0.1$, respectively.} 
}
\label{N_IT}
\end{figure}

\noindent According to ref.~\cite{ronchetti}, \textit{Respiratory viruses remain quiet for months, inactive but viable, within living cells. Then suddenly they activate and become virulent as they say, the infectious capacity grows to a maximum, after which it decreases. So, there is no valid reason to think that this Coronavirus behaves differently from others infections by SARS.}~\cite{ronchetti}. The present work starts from the following hypothesis: \textit{the SARS-CoV-2 behaves like the other viruses that cause respiratory diseases}. The common cold and influenza, do not confer any long-lasting immunity. Such infections do not give immunity upon recovery from infection, and individuals become susceptible again. Hence, according to the above-cited hypothesis we propose the following simplest compartmental model:
\begin{equation}\label{D0}
S +I \xrightarrow{\sigma} 2I
\end{equation}
\noindent In our model the SARS-CoV-2 infection does not leave any immunity, thus individuals return back into the $S$ compartment. Hence, infectious people, after recovery, return back to the compartmental $S$. This added detail can be shown by including an $R$ class in the middle of the model
\begin{align}\label{D1}
&S +I \xrightarrow{\sigma} 2I\\
&I \xrightarrow{\gamma} R\xrightarrow{\gamma_1} S\nonumber
\end{align}
\noindent From scheme~(\ref{D1}), we get the O.D.E.s for $S$, $R$, and $I$: 
\begin{align}\label{D2I}
&\frac{dS}{dt}=-\sigma\frac{S}{N_{Tot.}} I+\gamma_ 1R\\
&\frac{dI}{dt}=\sigma\frac{S}{N_{Tot.}} I-\gamma I\nonumber\\
&\frac{dR}{dt}=\gamma I-\gamma_1 R\nonumber
\end{align}
\noindent with $N_{Tot.}$ denoting the total population. By assuming that the dynamics of $R$ is much faster that those of $S$ and $I$, we may set $d R/dt\simeq 0$ and system~(\ref{D2I}) reduces to
\begin{align}\label{D2}
&\frac{dS}{dt}\simeq-\sigma\frac{S}{N_{Tot.}} I+\gamma I\\
&\frac{dI}{dt}=\sigma\frac{S}{N_{Tot.}} I-\gamma I\nonumber
\end{align} 
\noindent which corresponds to the model
\begin{align}\label{D2a}
&S +I \xrightarrow{\sigma} 2I\\
&I \xrightarrow{\gamma} S\nonumber
\end{align}
\noindent In literature, the model~(\ref{D2a}) is referred to as the $SIS$-\textit{model} (see, for example, \cite{SIS}). From Eq.~(\ref{D2}) we get the conservation relation
\begin{equation}\label{D3}
\frac{dS}{dt}+\frac{dI}{dt}=0\qquad{\rm or}\quad S+I=N_{Tot.}
\end{equation} 
\noindent Hence, the dynamics of infectious is governing by the logistic model
\begin{equation}\label{D4}
\frac{dI}{dt}=(\sigma-\gamma)I\left(1-\frac{\sigma}{N_{Tot.}(\sigma-\gamma)}I\right)
\end{equation} 
\noindent or
\begin{align}\label{D5}
&\frac{dI}{dt}={\tilde\alpha} I\left(1-\frac{I}{K}\right)\qquad{\rm with}\\
&{\tilde\alpha}\equiv\sigma\left(1-\frac{\gamma}{\sigma}\right)\quad;\quad K\equiv N_{Tot.}\left(1-\frac{\gamma}{\sigma}\right)\nonumber
\end{align} 
\noindent where $\tilde\alpha$ and $K$ denote the \textit{linear growing rate of the COVID-19} and the \textit{carrying capacity}, respectively. 
\noindent The lockdown measures are mainly based on the isolation of the susceptible individuals, eventually with the removal of infected people by hospitalisation. In our model, the effect of the lockdown measures are taken into account by introducing in the $SIS$-model the \textit{lockdown-induced decrease rate} $c(t)$
\begin{align}\label{D6}
&S +I \xrightarrow{\sigma} 2I\\
&I \xrightarrow{\gamma+c(t)} S \nonumber
\end{align}
\noindent where
\begin{align}\label{D7}
&c(t_L)=0\qquad {\rm for} \quad t=t_L\\
&c(t)>0\qquad\ \ {\rm for} \quad t>t_L\nonumber
\end{align}
\noindent with $t_L$ denotes the time when the lockdown measures are applied. The corresponding deterministic differential equations for the COVID-19 model in presence of the lockdown measures reads then:
\begin{align}\label{D8}
&\frac{dS}{dt}=-\sigma\frac{S}{N_{Tot.}} I+\gamma I+c(t)I\\
&\frac{dI}{dt}=\sigma\frac{S}{N_{Tot.}} I-\gamma I-c(t)I\nonumber
\end{align} 
\noindent Scheme~(\ref{D6}) may be referred to as the $(SIS)_L$-\textit{model} where $L$ stands for $Lockdown$. The general expression for the lockdown-contribution, $c(t)$, may be cast into the form\footnote{Here, we shall not discuss on the completeness of the basis functions $\{ {\hat t}^{i+1}-{\hat t}^{-j}\}_{i,j=0,1,\dots}$.}
\begin{equation}\label{D9}
c({\hat t})=\sum_{i,j=0}^\infty\beta_{ij}\left({\hat t}^{i+1}-{\hat t}^{-j}\right)\qquad{\rm with}\quad {\hat t}\geq 1
\end{equation} 
\noindent with ${\hat t}$ denoting the \textit{normalised time} ${\hat t}\equiv t/t_L$ and $\beta_{ij}$ are real numbers, subject to the condition $c({\hat t})>0$ for ${\hat t}>1$, respectively. Notice that expression~(\ref{D9}) satisfies condition~(\ref{D7}) and is able to dampen the growing trend of the number of infectious. We are looking for an expression for the lockdown, which grows at most linearly in time. So, the relevant terms in expression~(\ref{D9}) are
\begin{equation}\label{D9a}
c({\hat t})\simeq \beta_{00}({\hat t}-1) +\beta_{01} \left(\frac{{\hat t}^2-1}{\hat t}\right)\qquad{\rm with}\quad {\hat t}\geq 1
\end{equation} 
\noindent However, fittings with the real data correspond to $\beta_{00}\simeq 0$ (ref. to \cite{sonnino}, \cite{sonnino1}, and \cite{mathus}). Finally, we get
\begin{equation}\label{D10}
c({\hat t})\simeq \beta\left(\frac{{\hat t}^2-1}{\hat t}\right)\qquad{\rm with}\quad\beta>0\quad {\rm and}\quad {\hat t}\geq 1
\end{equation}
\noindent with $\beta\equiv \beta_{01}=const.$ denoting the \textit{intensity of the lockdown measures}. Eq.~(\ref{D10}) is the expression for the lockdown measures that will be considered in this work. Notice that Eq.~(\ref{D10}) generalises the lockdown term introduced in ref.~\cite{sonnino}. By taking into account Eq.~(\ref{D3}), the deterministic differential equation for the infectious people, in presence of the lockdown measures, reads
\begin{equation}\label{D11}
\frac{dI}{d{\hat t}}=\alpha I\left(1-\frac{I}{K}\right)-\beta\left(\frac{{\hat t}^2-1}{\hat t}\right) I
\end{equation}
\noindent with $\alpha\equiv{\tilde\alpha} t_L$. As shown in the Appendix, the exact solution of Eq.~(\ref{D11}) reads
\begin{equation}\label{D12}
I=\frac{A{\hat t}^\beta\exp(-({\hat t}-\alpha/\beta)^2/\sigma)}{1+A\alpha/K\int_1^{\hat{t}}x^\beta\exp(-(x-\alpha/\beta)^2/\sigma)dx}
\end{equation} 
\noindent where $\sigma\equiv 2/\beta$. Expression~(\ref{D12}) can be brought into the form
\begin{equation}\label{D13}
I=\frac{I_0{\hat t}^\beta\exp((1-\alpha/\beta)^2/\sigma)\exp(-({\hat t}-\alpha/\beta)^2/\sigma)}{1+(I_0\alpha/K)\exp((1-\alpha/\beta)^2/\sigma)\int_1^{\hat{t}}x^\beta\exp(-(x-\alpha/\beta)^2/\sigma)dx}
\end{equation} 
\noindent with $I_0$ denoting the value of the total cases at the time when the lockdown measures are applied, i.e. $I_{{\hat t}=1}=I_0$. Eq.~(\ref{D13}) can be cast into the form
\begin{equation}\label{D14}
I\simeq I_0{\hat t}^\beta\exp((1-\alpha/\beta)^2/\sigma)\exp(-({\hat t}-\alpha/\beta)^2/\sigma)
\end{equation}
\noindent Fig.(\ref{N_IT}) shows two solutions of the model~(\ref{D11}) for Italy, during the first wave of infection by SARS-CoV-2. The red dotted line refers to the solution without the lockdown measures and the dark dotted line to the solution when the lockdown measures were applied.

\subsection{\bf The Second Wave of the COVID-19 Pandemic in the $(SIS)_L$-model}\label{SW}
\noindent The \textit{basic reproduction number}, $R_0$, is defined as the expected number of secondary cases produced by a single (typical) infection in a completely susceptible population. An epidemic occurs if the number of infected individuals increases, i.e., $dI/dt > 0$
\begin{equation}\label{et1}
\sigma\frac{S}{N_{Tot.}} I-\gamma I>0\qquad{\rm or}\quad \frac{\sigma}{\gamma}\frac{S}{N_{Tot.}}>1
\end{equation} 
\noindent In this case, $R_0=\sigma/\gamma$. When the lockdown measures are interrupted, the dynamics of the process are governed again by Eq.~(\ref{D5}). For Italy, in the first wave we had $R_0 >1$, so $K>0$. From Eq.~(\ref{D5}) we get
\begin{equation}\label{et2}
\frac{\sigma}{\gamma} >1\qquad {\rm hence}\quad\alpha>0\qquad{\rm and}\quad \lim_{{\hat t}\rightarrow\infty}I=K
\end{equation} The above relations biologically mean that the disease will remain permanently endemic in the population. Notice that the inflection point of the logistic curve~(\ref {D5}) occurs at time $t=t_f$ with
\begin{equation}\label{et3}
{\hat t}_f=\frac{1}{\alpha}\log\left(\frac{K}{I_L}-1\right)
\end{equation} 
\noindent with $I_L$ denoting the number of the infectious people when the lockdown measures start to be applied. It is easily checked that the inflection point, $P_f$, has coordinates $P_f=({\hat t}_f,I_f=K/2)$. In our forthcoming reasoning, we shall see that the inflection point plays an important role in our analysis.  To summarize, after removing the lockdown measurements, say at time ${\hat t}_1$, the number of infectious people will start to rise again from the value $I_1$, with $I_1\equiv I_{t={\hat t}_1}$, following again a growth rate given by the logistic curve~(\ref {D5}). This new growth of outbreak is referred to as the \textit{second wave of COVID-19}. Now, our goal is to answer the following question: "\textit{Given a fixed value of the infectious people, say ${\tilde I}$, determine the minimum value that $I_1$ should have so that the number of infectious people is less than ${\hat I}$, during the entire period from ${\hat t}_1$ to ${\hat t}_f$}". We note that within the interval ${\hat t}_1\leq {\hat t}\leq {\hat t}_f$ the number of infectious people grows quite slowly. Let us indicate with ${\hat t}_2$ the time when it is decided to interrupt the lockdown measures. Hence, during the time interval $\Delta {\hat t}={\hat t}_2-{\hat t}_1$, population is no longer subject to the lockdown measures. To answer the above-mentioned question we need firstly to find the range of values that can be assumed by $I_1$ so that in the interval $\Delta {\hat t} < {\hat t}_f-{\hat t}_1$, the number of infectious individuals is lower than the ordinate of the inflection point, i.e. $I_1<I_f=K/2$ for ${\hat t}\in\Delta {\hat t}$. It is easily checked that $I_1$ should be within the range 
\begin{equation}\label{et4}
0\leq I_1 <\frac{K}{1+\exp (\alpha\Delta {\hat t})}
\end{equation} 
\noindent Now, we have to require that after a lapse of time $\Delta {\hat t}$, the number of the infectious individuals must be lower than a value ${\tilde I}$ fixed \textit{a priori}, with ${\tilde I}<K/2$. The result is easily obtained by writing the initial condition as
\begin{equation}\label{et5}
I_1 =\frac{K}{1+\exp (\alpha\Delta {\hat t})}-Kx\qquad{\rm with}\quad 0<x\leq \frac{1}{1+\exp (\alpha\Delta {\hat t})}
\end{equation} 
\noindent and by determining the range of values of $x$ such that the inequality 
\begin{equation}\label{et6}
\frac{K}{1+(K/I_1-1)\exp(-\alpha\Delta {\hat t})}\leq {\tilde I}<\frac{K}{2}
\end{equation} 
\noindent is satisfied. After a little algebra, we get
\begin{equation}\label{et6}
0\leq I_1\leq {\tilde I}\frac{2+\exp{\alpha\Delta {\hat t}}+\exp(-\alpha\Delta {\hat t})}{(1+\exp(\alpha\Delta {\hat t})({\tilde I}/K(\exp(\alpha\Delta {\hat t})+\exp(-\alpha\Delta {\hat t})+1+\exp(\alpha\Delta {\hat t}))}
\end{equation} 

\begin{figure}[hbt!]
\hskip 1.3truecm
\includegraphics[width=9.7cm, height=6cm]{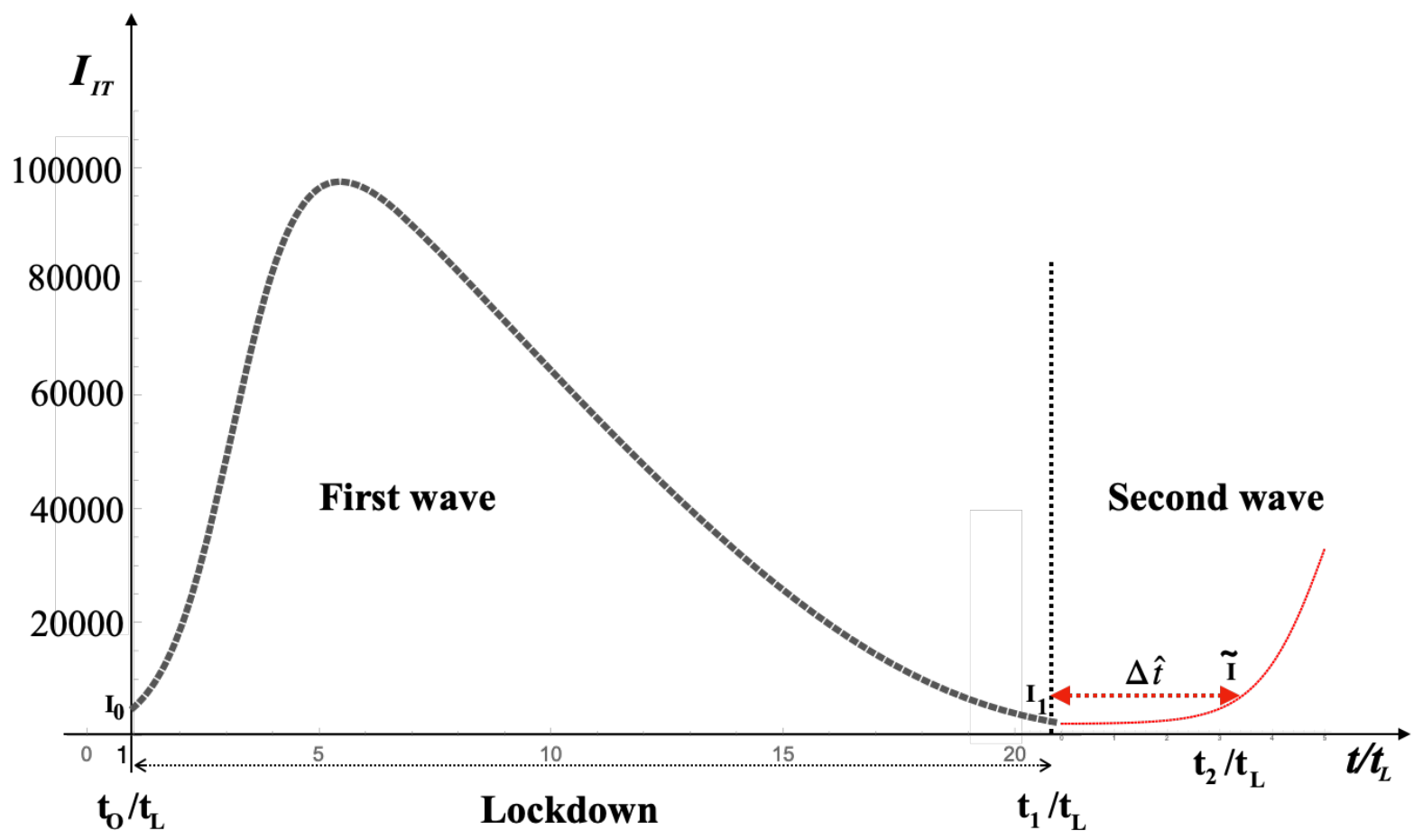}
\caption{
\textit{{\bf Schematic representation of the second wave of SARS-CoV-2 infection.} The Lockdown measures are applied during the period $ {{\hat t} _0}-{{\hat t} _1}$. At the end of the confinement measures, the number of infectious dropped to $I_1$. During the time interval $\Delta{\hat t}={{\hat t}_2}-{{\hat t} _1}$, the lockdown measures are not applied and the infection by SARS-CoV-2 (starts growing again (\textit {second wave of Coronavirus infection}) until the preset value ${\tilde I}$ is reached, which corresponds to $I({\hat t}_2)$. During the whole time interval $\Delta {\hat t}$, the number of infectious individuals still stay below a set threshold ${\tilde I}$ if $I_1$ satisfies condition~(\ref{et6}).} 
}
\label{SecWave1}
\end{figure}

\noindent To summarise, if the lockdown measures are interrupted at the time ${\hat t}_1$, during the successive period of time $\Delta {\hat t}={\hat t}_2-{\hat t}_1$ the number of infectious remains inferior to a preset number ${\tilde I}$ if the number of the infectious at the end of the lockdown measures is inferior to $I_1$, with $I_1$ given by the r.h.s. of Eq.~(\ref{et6}). A Schematic representation of the second wave of SARS-CoV-2 infection is shown in Fig.~\ref{SecWave1}.

\begin{figure*}[htb]
  \hfill
  \begin{minipage}[t]{.48\textwidth}
    \centering
    \includegraphics[width=5cm,height=5cm]{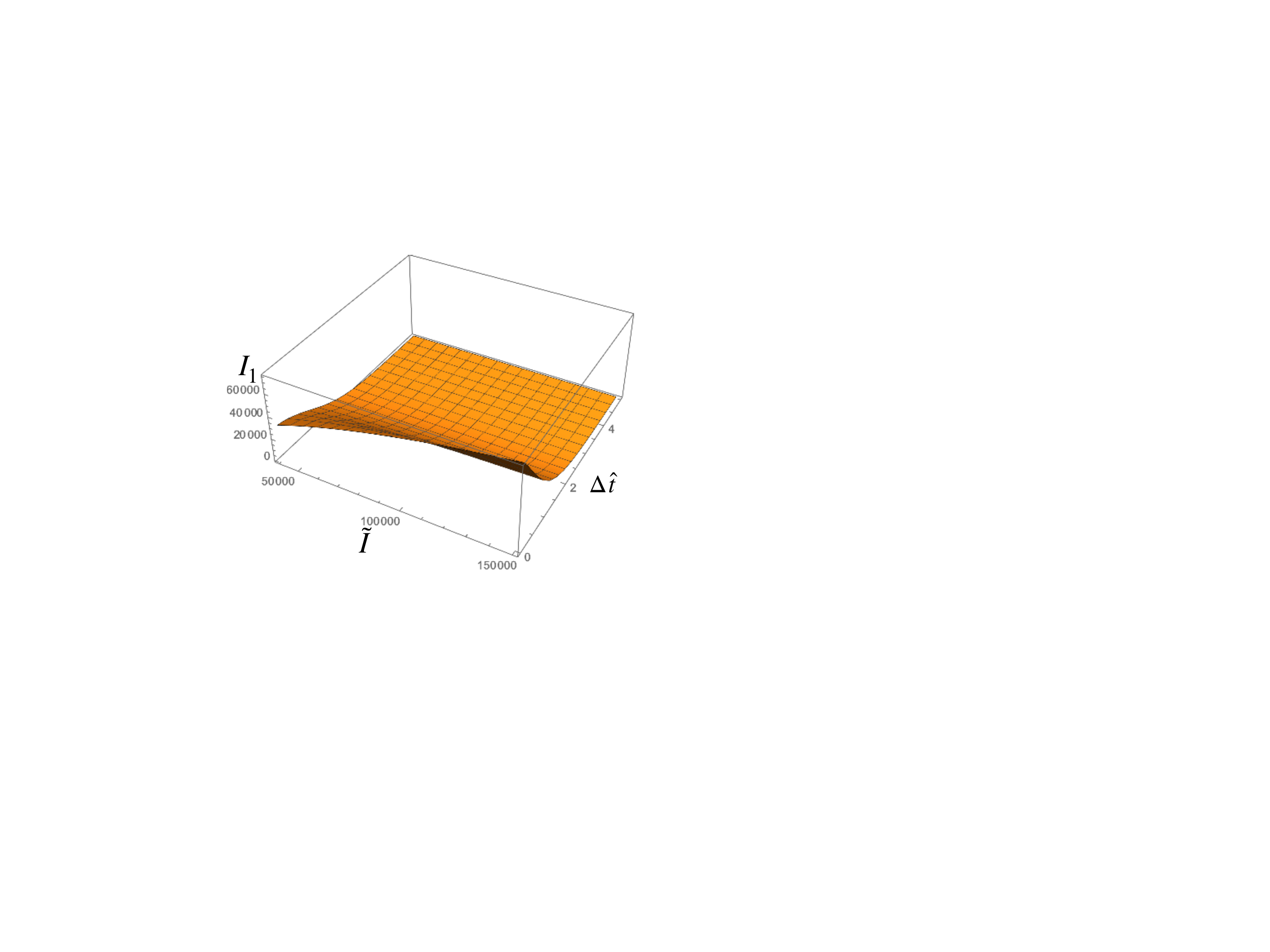}
    \caption{\textit{$I_1$ against $\tilde I$ and $\Delta{\hat t}$ in the range $50000\leq{\tilde I}\leq K_{IT}$.}}
    \label{SecWave2}
  \end{minipage}
  \hfill
    \begin{minipage}[t]{.48\textwidth}
      \centering
      \includegraphics[width=5cm,height=5cm]{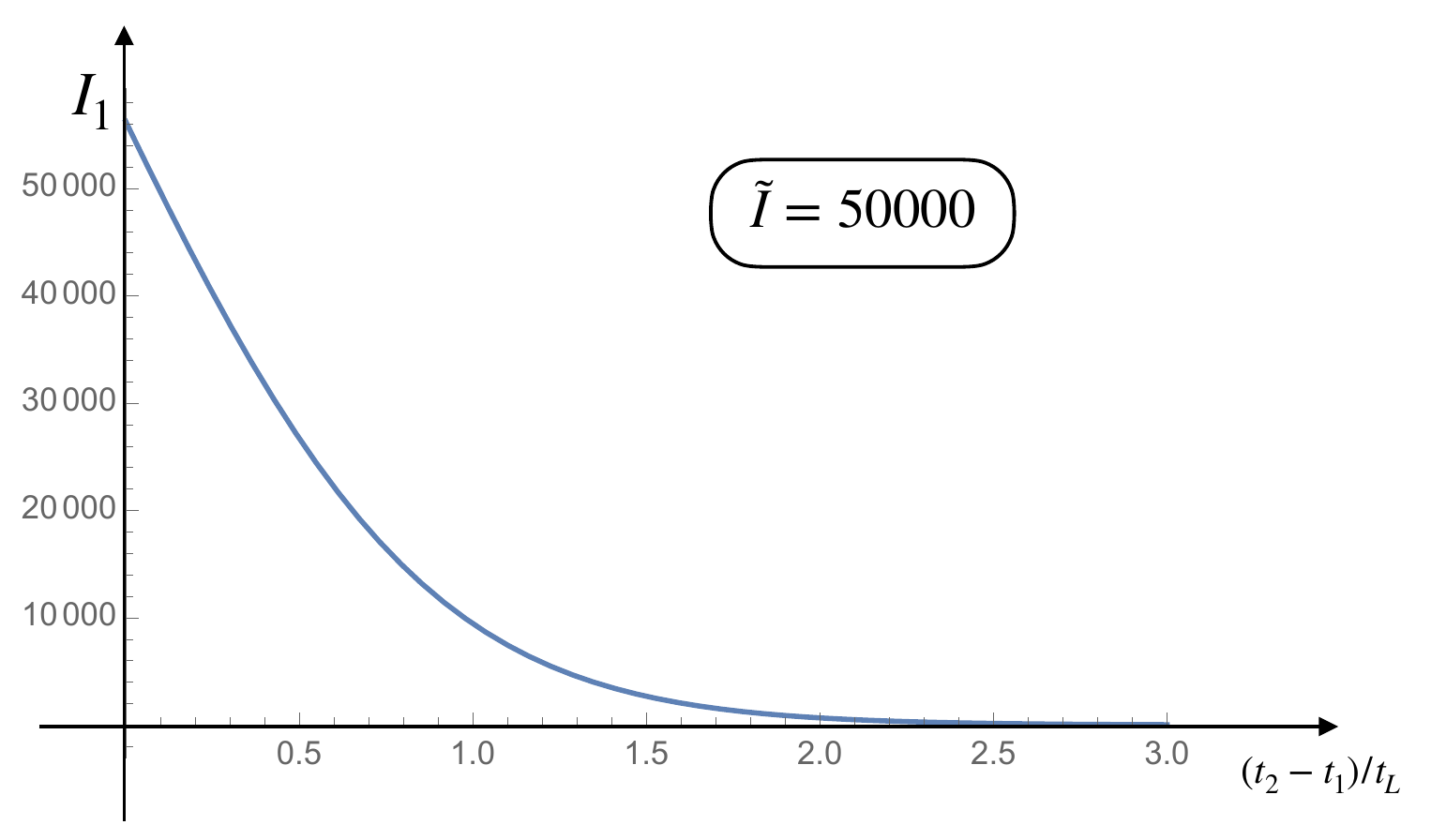}
      \caption{\textit{As shown, if we set the value of ${\tilde I} = 50000$, the maximum value of $I_1$ is $I_1= 9500$ for $t=t_L$, and $I_1=700$ for $t=2t_L$, respectively.}}
      \label{SecWave3}
    \end{minipage}
  \hfill
\end{figure*}

\noindent  As an example, let us consider the case of the Italian situation. Fig.~\ref{SecWave2}. reports $I_1$ against ${\tilde I}$ and $\Delta{\hat t}$ in the range $50000\leq{\tilde I}\leq K_{IT}$. As shown in Fig.~\ref{SecWave3}., if we set the value of ${\tilde I} = 50000$, the maximum value of $I_1$ is $I_1\simeq 9500$ for $t\simeq t_L$, and $I_1\simeq 700$ for $t\simeq 2t_L$, respectively.

\section{The Stochastic $(SIS)_L$-Model}\label{SM}
\begin{figure}[hbt!]
\hskip 0.5truecm
\includegraphics[width=11cm, height=7cm]{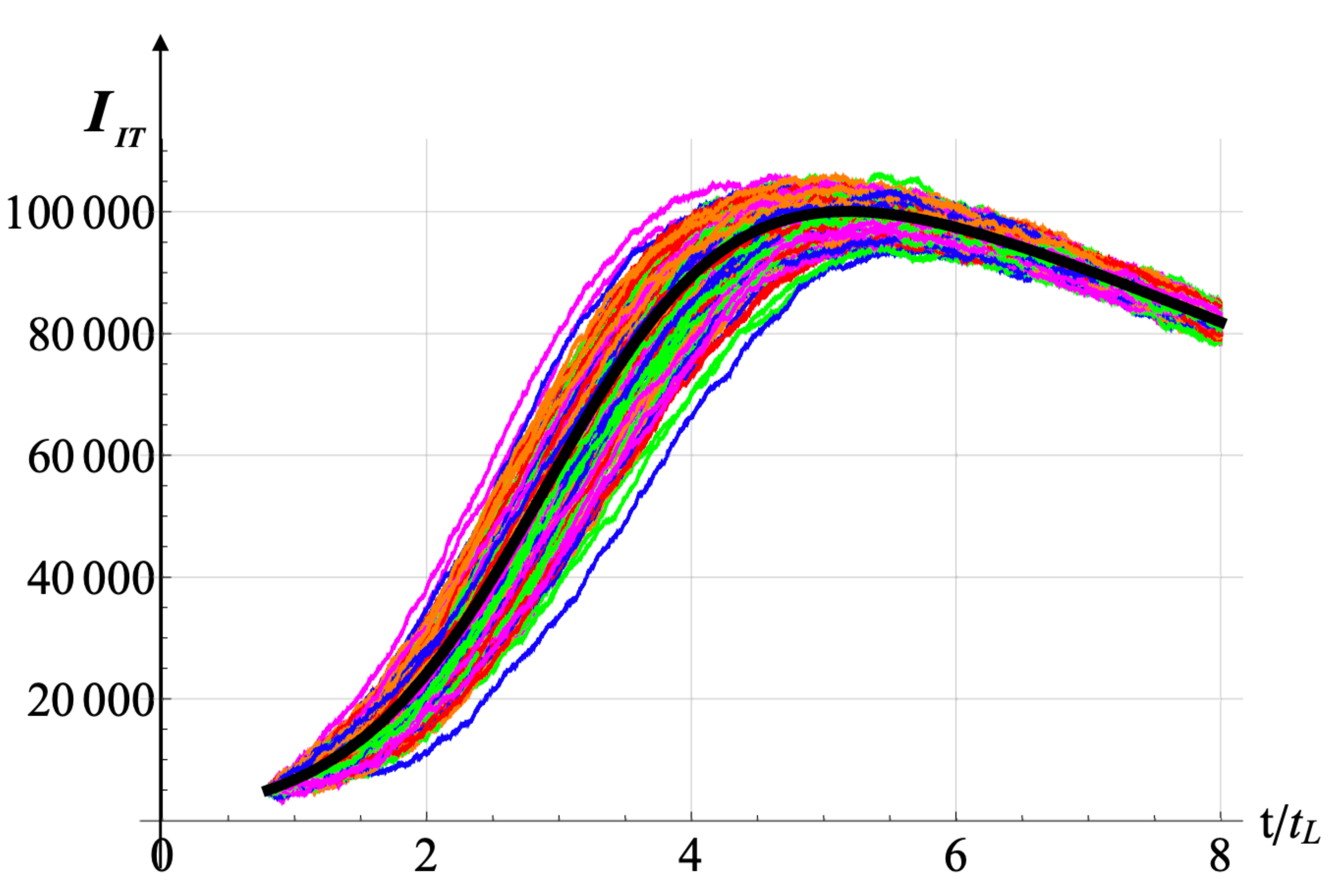}
\caption{
\textit{{\bf Solution of the stochastic equation for Italy.} Solutions of Eqs~(\ref{S1}) and (\ref{S2}) for Italy - first wave of infection by SARS-CoV-2 - with $\eta=0.05$ and for 200 realisations.  The values of the other parameters are $I_0$=5000, $K=150000$, $\alpha=1.5$, and $\beta=0.1$, respectively. The black thick curve is the numerical solution of the deterministic equation~(\ref{D11}). The black curve is the solution of the deterministic equation.} 
}
\label{I_Stoch_IT1}
\end{figure}

\noindent The Stochastic version of Eq.~(\ref{D11}) reads
\begin{equation}\label{S1}
\frac{dI}{d{\hat t}}=\alpha I\left(1-\frac{I}{K}\right)-\beta\left(\frac{{\hat t}^2-1}{{\hat t}}\right)I+\ \xi({\hat t})
\end{equation}
\noindent with $\xi({\hat t})$ denoting a white noise
\begin{align}\label{S2}
&<\xi({\hat t})>=0\\
&<\xi({\hat t})\xi({\hat t}')>=\eta\delta({\hat t}-{\hat t}')\nonumber
\end{align}
\noindent and $\sqrt\eta$ is the intensity of the noise. $\delta$ denotes the Dirac delta function (distribution). Let us now consider the system at the reference state $I_{RS}$, which is the solution of the deterministic equation 
\begin{equation}\label{S3}
\frac{dI_{RS}}{d{\hat t}}=\alpha I_{RS}\left(1-\frac{I_{RS}}{K}\right)-\beta\left(\frac{{\hat t}^2-1}{\hat t}\right)I_{RS}
\end{equation}
\noindent subject to a perturbation of small amplitude $\delta I$, i.e.
\begin{equation}\label{S4}
I(t)=I_{RS}(t)+\delta I(t)
\end{equation}
\noindent Our goal is to compute the relevant statistical correlation functions of this processes i.e. $<\delta I({\hat t})\xi({\hat t})>$ and $<\delta I(t)\delta I(t')>$. Our task can easily be accomplished by recalling the following theorem~(see, for example, \cite{malek}, \cite{malek1}, \cite{gardiner}, \cite{vankampen}):

\noindent \textit{For a system of Langevin equations describing the temporal evolution of the processes} $(x_1,\dots, x_n)$:
\begin{align}\label{S5}
&\frac{dx_i}{dt}=f_i(x_1,\dots ,x_n)+\xi_i(t)\qquad i=1,\dots, n\\
&<\xi_i(t)>=0\quad;\quad <\xi_i(t)\xi_j(t')>=\eta_{ij}\delta (t-t')\nonumber
\end{align}
\noindent \textit{the following equations are valid}:
\begin{align}\label{S6}
&<x_i(t)\xi_j(t')>=0\qquad{\rm if}\quad t'>t\\
&<x_i(t)\xi_j(t)>=\frac{1}{2}\eta_{ij}\nonumber 
\end{align}
\noindent Now, we assume that the perturbation $\delta I$ is of order $\sqrt\eta$ (considered as a small parameter) and we develop Eq.~(\ref{S1}) up to the first order in $\sqrt\eta$. We get
\begin{align}\label{S7}
&\frac{dI_{RS}({\hat t})}{d{\hat t}}=\alpha I_{RS}({\hat t})\left(1-\frac{I_{RS}({\hat t})}{K}\right)-\beta\left(\frac{{\hat t}^2-1}{\hat t}\right)I_{RS}({\hat t})\ {\rm (at\ the\ zero\ order)}\nonumber\\
&\frac{d}{d{\hat t}}\delta I({\hat t})=\left(\alpha-\frac{2\alpha}{K}I_{RS}({\hat t})-\beta\frac{{\hat t}^2-1}{{\hat t}}\right)\delta I({\hat t})+\xi({\hat t})\ \ {\rm (at\ the\ first\ order)}
\end{align}
\noindent Solution $I_{RS}({\hat t})$ is given by Eq.~(\ref{D13}) and the exact solution for $\delta I({\hat t})$ is easily obtained:
\begin{align}\label{S8}
I_{RS}({\hat t})=&\frac{I_{0RS}{\hat t}^\beta\exp((1-\alpha/\beta)^2/\sigma)\exp(-({\hat t}-\alpha/\beta)^2/\sigma)}{1+(I_{0RS}\alpha/K)\exp((1-\alpha/\beta)^2/\sigma)\int_1^{\hat{t}}x^\beta\exp(-(x-\alpha/\beta)^2/\sigma)dx}\nonumber\\
\delta I({\hat t})=&t^\beta\exp\left(\alpha({\hat t}-1)-1/2\beta({\hat t}^2-1)-2\alpha G({\hat t})/K\right)\Big(\delta I(1)\nonumber\\
&+\int_1^{\hat t}dx\ \xi(x)\left(x^{-\beta}\exp(-\alpha (x-1)+1/2\beta (x^2-1)+2\alpha G(x)/K\right)\nonumber\\
&{\rm with}\qquad G({\hat t})\equiv\int_1^{\hat t}I_{RS}(x)dx
\end{align}
\noindent Solution~(\ref{S8}) allows to check the validity of Eq.~(\ref{S6}). Indeed, the correlation function $\delta I(t)$-\textit{noise} at the same time reads
\begin{equation}\label{S9}
<\delta I({\hat t})\xi({\hat t})>=\frac{\eta}{2}
\end{equation}
\noindent where Eqs~(\ref{S5}) and the properties of the Dirac's delta have been taken into account. The equation for the second moment $<\delta I^2({\hat t})>$ is easily obtained by multiplying both sides of the second equation of the system~(\ref{S7}) by $2\delta I(t)$ and by taking into account Eq~(\ref{S6}). We get 
\begin{equation}\label{S10}
\frac{d}{d{\hat t}}<\left(\delta I({\hat t}\right))^2>=2\left(\alpha-\frac{2\alpha}{K}I_{RS}({\hat t})-\beta\frac{{\hat t}^2-1}{{\hat t}}\right)<\left(\delta I({\hat t})\right)^2>+\eta
\end{equation}
\noindent The solution of Eq.~(\ref{S10}) provides the expression for the second moment
\begin{align}\label{S11}
<\left(\delta I({\hat t}\right))^2>=&\left({\hat t}^{2\beta}\exp(2\alpha({\hat t}-1)-\beta({\hat t}^2-1)-4\alpha G({\hat t})/K\right)\!\!\Big(<\!\left(\delta I(1)\right)^2\!>\nonumber\\
&\!+\eta \int_1^{\hat t}x^{-2\beta}\exp(-2\alpha (x-1)+\beta (x^2-1)+4\alpha G(x)/K)\!dx\Bigl)
\end{align}
\noindent From solution~(\ref{S8}) we also get the correlation function for $\delta I $ 
\begin{equation}\label{S12}
<\delta I(1)\delta I({\hat t})> = {\hat t}^\beta\exp(\alpha({\hat t}\!-\!1)\!-\!\beta/2({\hat t}^2\!-\!1)\!-\!2\alpha G({\hat t})/K)\!<(\delta I(1))^2>
\end{equation}
\noindent 
\noindent Fig.~\ref{I_Stoch_IT1}. shows the simulation of 200 trajectories of Eq.~(\ref{S1}) for Italy using the Order-2 Stochastic Runge-Kutta integration method. The thick black curve is the numerical solution of the deterministic equation~(\ref{D11}). The intensity of the noise is $\eta=0.05$ and the values of other parameters are reported in the figure caption of Fig.~\ref{I_Stoch_IT1}.
\noindent Figs.~({\ref{CFII}) and ({\ref{CFI0I}) show the correlation functions $<(\delta I({\hat t}))^2>$ and $<\delta I(1)\delta({\hat t})>$ for Italy, first wave of infection by SARS-CoV-2. The values of the parameters are reported in the figure captions.
\begin{figure*}[htb]
  \hfill
  \begin{minipage}[t]{.45\textwidth}
    \centering
    \includegraphics[width=5cm,height=5cm]{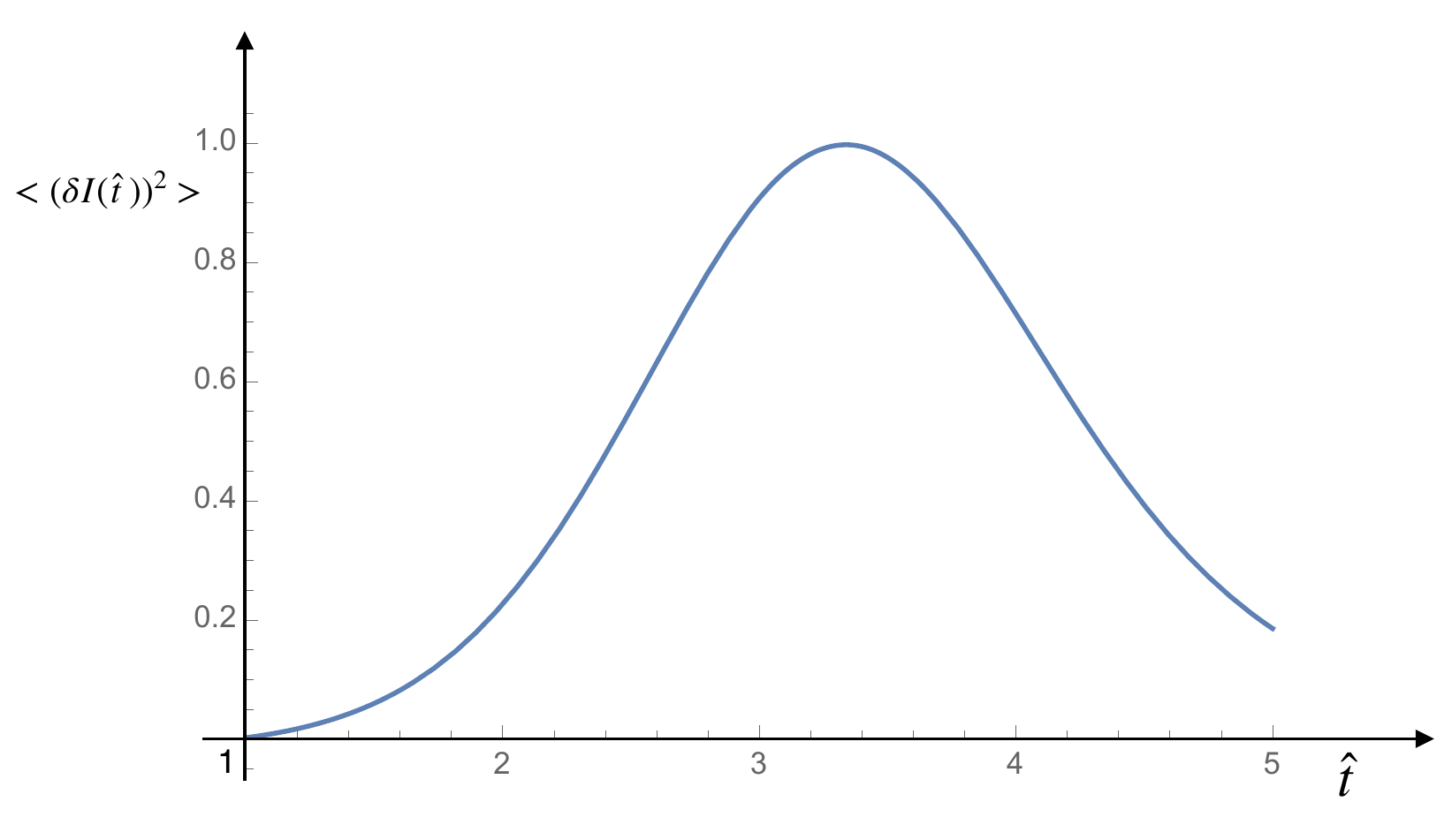}
    \caption{
    \textit{{\bf Correlation Function} $<(\delta I({\hat t}))^2>$ {\bf for Italy, first wave of infection by SARS-CoV-2.} This correlation function corresponds to the solution of Eq.~(\ref{S10}). The values of the parameters are $I_0$=5000, $K=150000$, $\alpha=1.5$, and $\beta=0.1$, respectively.} }
    \label{CFII}
  \end{minipage}
  \hfill
    \begin{minipage}[t]{.45\textwidth}
      \centering
      \includegraphics[width=5cm,height=5cm]{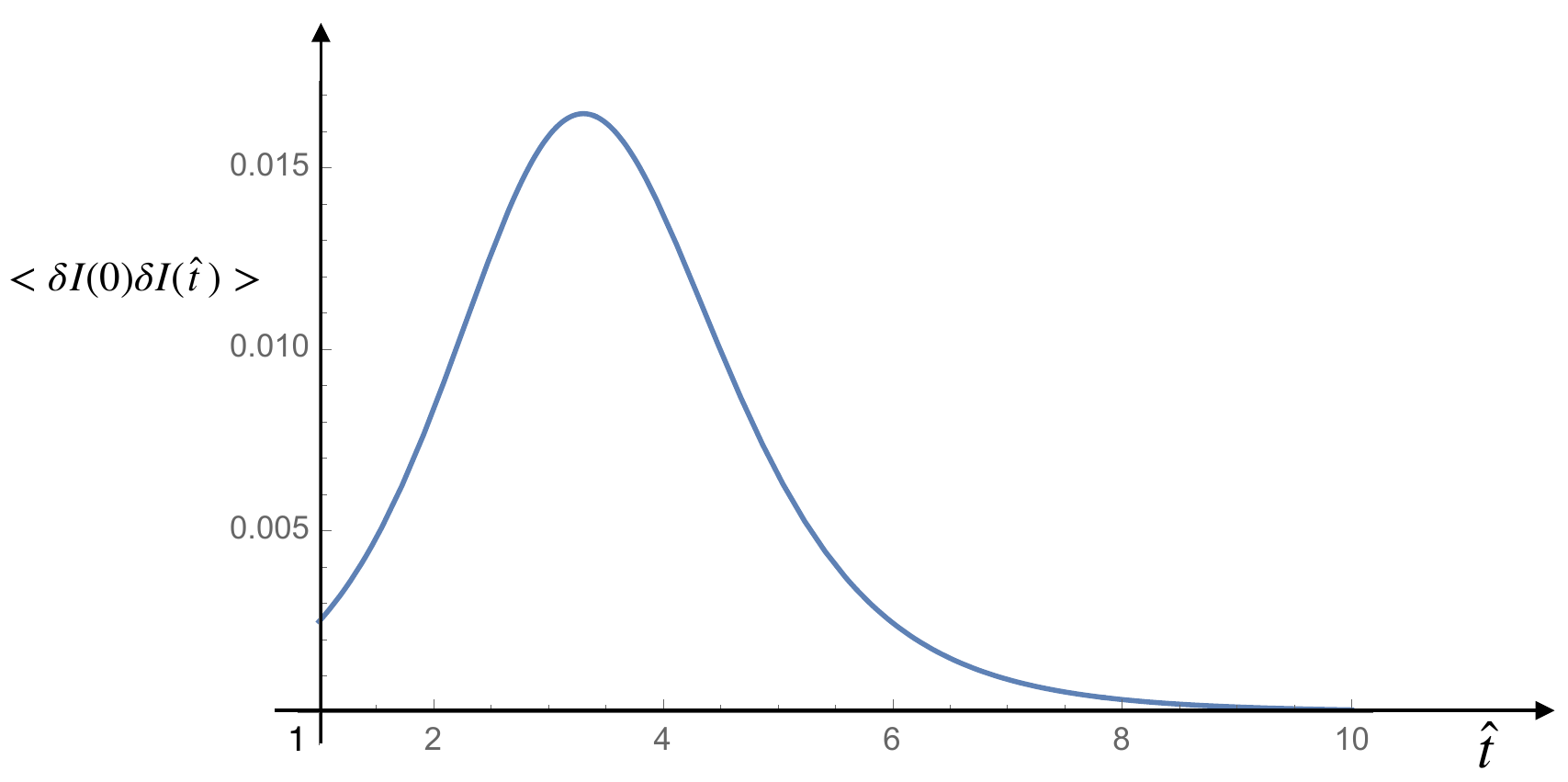}
      \caption{
\textit{{\bf Correlation Function for the Infectious people} $<\delta I(1)\delta({\hat t})>$. Plot of Eq.~(\ref{S12}) with $I_0$=5000, $K=150000$, $\alpha=1.5$, and $\beta=0.1$, respectively.}
      }
      \label{CFI0I}
    \end{minipage}
  \hfill
\end{figure*}
\noindent The Fokker-Planck equation, associated to Eq.~(\ref{S1}), describing the time evolution of the probability density function (PDF), $p(I,t)$, to find the number of infectious people at time $\hat t$ reads \cite{risken}
\begin{align}\label{S13}
&\frac{\partial}{\partial {\hat t}}p(I,{\hat t})=-\frac{\partial}{\partial I}[\mu({\hat t})p(I,{\hat t})]+D\frac{\partial^2}{\partial I^2}p(I,{\hat t)}\\
&\mu({\hat t})=\alpha I\left(1-\frac{I}{K}\right)-\beta\left(\frac{{\hat t}^2-1}{\hat t}\right)\quad ; \quad D=\frac{\eta^2}{2}\nonumber
\end{align}
\noindent with $\mu$ denoting the \textit{drift coefficient} and $D$ the \textit{diffusion coefficient}, respectively. Eq.~(\ref{S13}) is subject to the initial condition
\begin{equation}\label{S13a}
p(I,0)=\delta(I-I_0)
\end{equation}
\noindent To get the PDF, we proceed as follows. First, we simulate paths or realisations of Eq.~(\ref{S1}) with the indicated parameters. The employed algorithm is an Order-2 Stochastic Runge-Kutta like scheme. Regarding the discretisation in the numerical solution of the Langevin Eq.~(\ref{S1}) we use 500 time steps to build each stochastic path. The PDF is computed numerically by constructing an histogram at each instant of time with a sufficiently number of realisations. The standard number used here is 22000 paths. The values of the parameters are $\alpha= 1.5$, $\beta= 0.1$, respectively, and for the PDF algorithm we use 800 bins in the construction of the corresponding histograms at the indicated times. The intensity of the noise, $\eta= 0.05$, has been estimated according to Eq.~(\ref{A11}) (see Appendix 1). Figures~(\ref{SMF1}) to (\ref{SMF8}) represent a set of screenshots at the indicated times ${\hat t}=t/t_L$.
\begin{figure*}[htb]
  \hfill
  \begin{minipage}[t]{.45\textwidth}
    \centering
    \includegraphics[width=5cm,height=5cm]{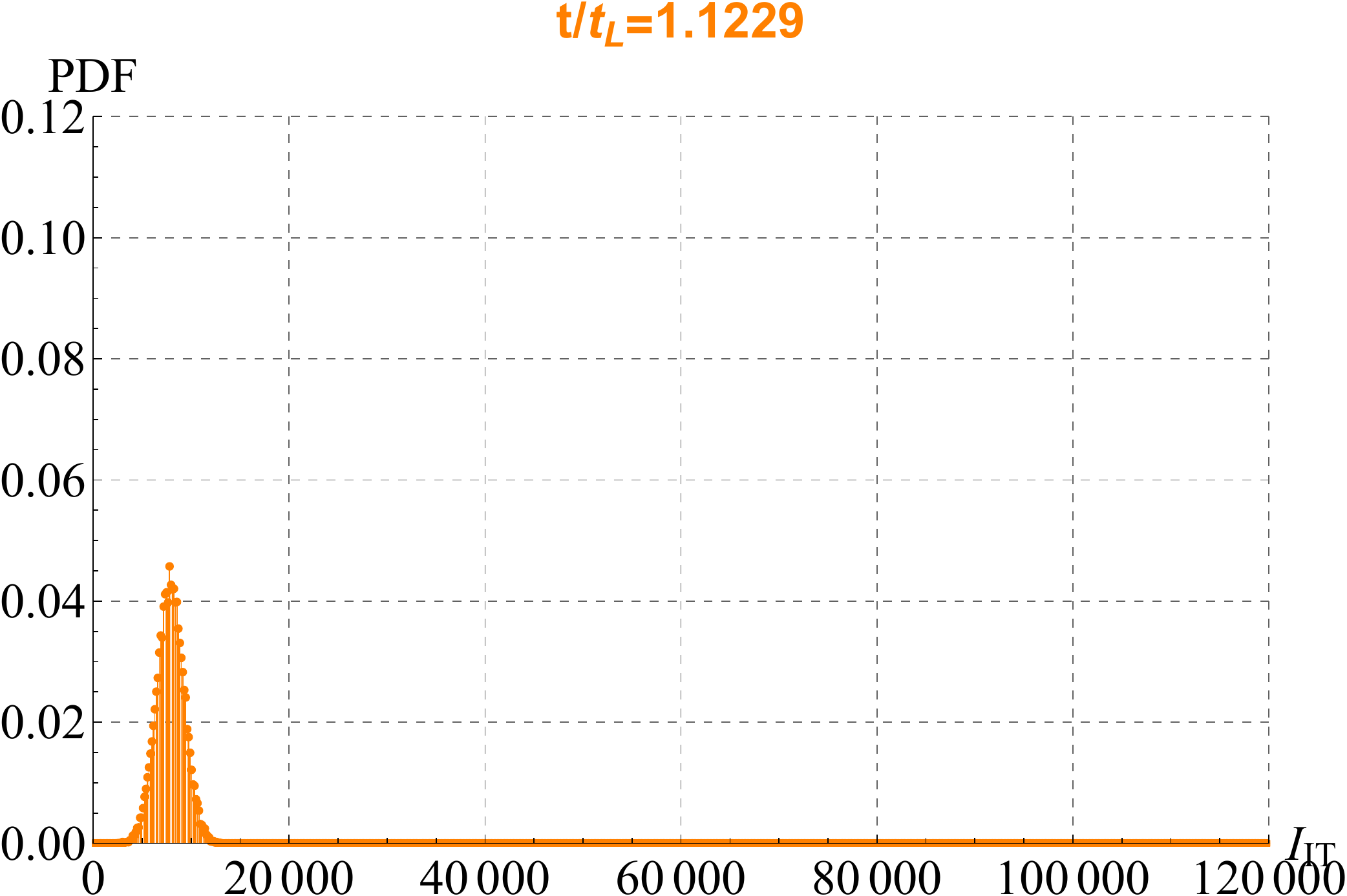}
    \caption{
    \textit{{\bf Probability density function (PDF), solution of Eq.~(\ref{S11}) for Italy, first wave of infection by SARS-CoV-2.} Screenshoot at time ${\hat t}=1.129$. The values of the parameters are $I_0$=5000, $K=150000$, $\alpha=1.5$, and $\beta=0.1$, respectively.} 
    }
    \label{SMF1}
  \end{minipage}
  \hfill
    \begin{minipage}[t]{.45\textwidth}
      \centering
      \includegraphics[width=5cm,height=5cm]{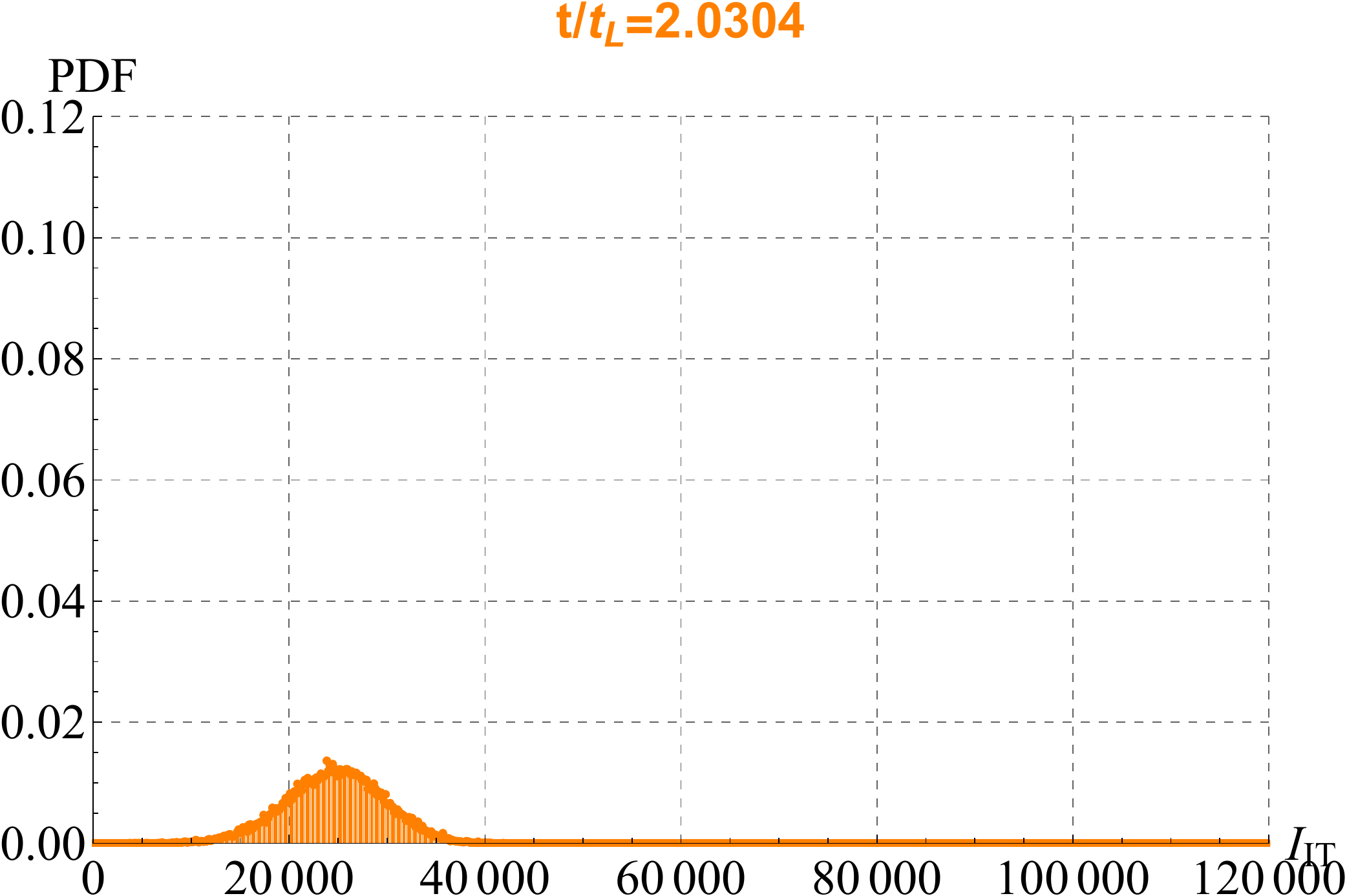}
      \caption{
\textit{{\bf Probability density function (PDF), solution of Eq.~(\ref{S11}).} Screenshoot at time ${\hat t}=2.0304$.} 
      }
      \label{SMF2}
    \end{minipage}
  \hfill
\end{figure*}
\begin{figure*}[htb]
  \hfill
  \begin{minipage}[t]{.45\textwidth}
    \centering
    \includegraphics[width=5cm,height=5cm]{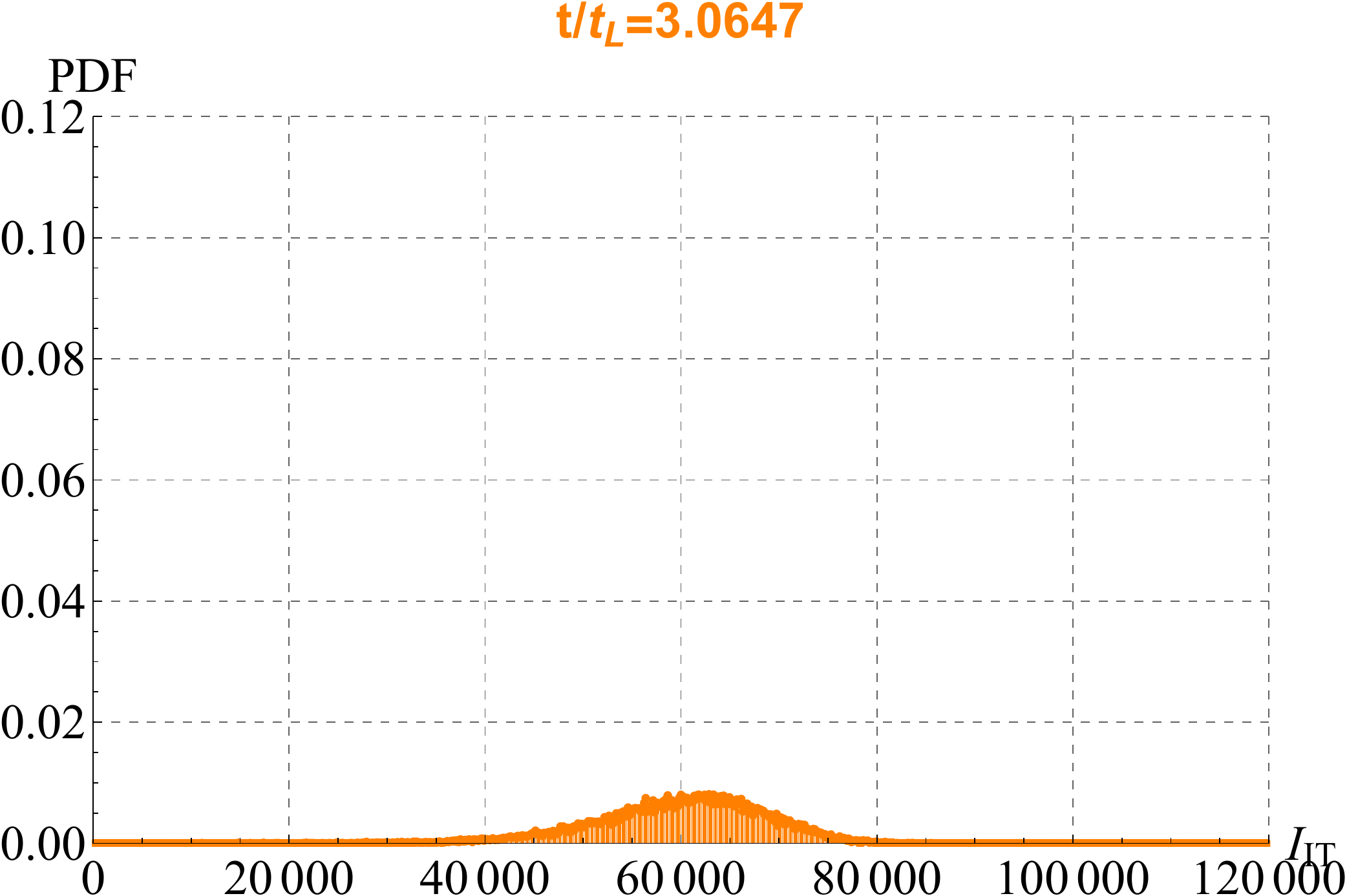}
    \caption{
    \textit{{\bf Probability density function (PDF), solution of Eq.~(\ref{S11}) for Italy, first wave of infection by SARS-CoV-2.} Screenshoot at time ${\hat t}=3.0647$.} 
    }
    \label{SMF3}
  \end{minipage}
  \hfill
    \begin{minipage}[t]{.45\textwidth}
      \centering
      \includegraphics[width=5cm,height=5cm]{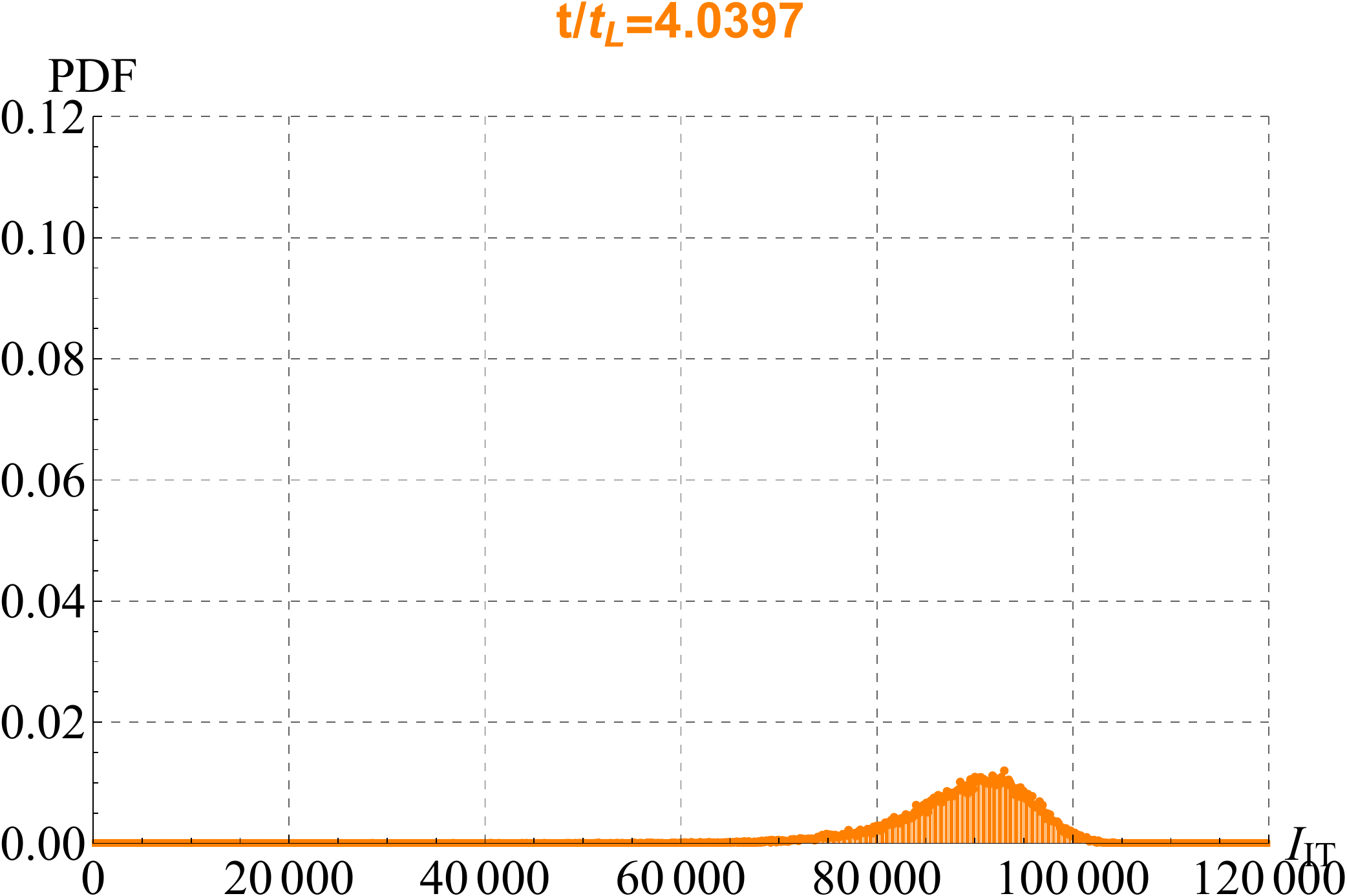}
      \caption{
\textit{{\bf Probability density function (PDF), solution of Eq.~(\ref{S11}).} Screenshoot at time ${\hat t}=4.0397$.} 
      }
      \label{SMF4}
    \end{minipage}
  \hfill
\end{figure*}
\begin{figure*}[htb]
  \hfill
  \begin{minipage}[t]{.45\textwidth}
    \centering
    \includegraphics[width=5cm,height=5cm]{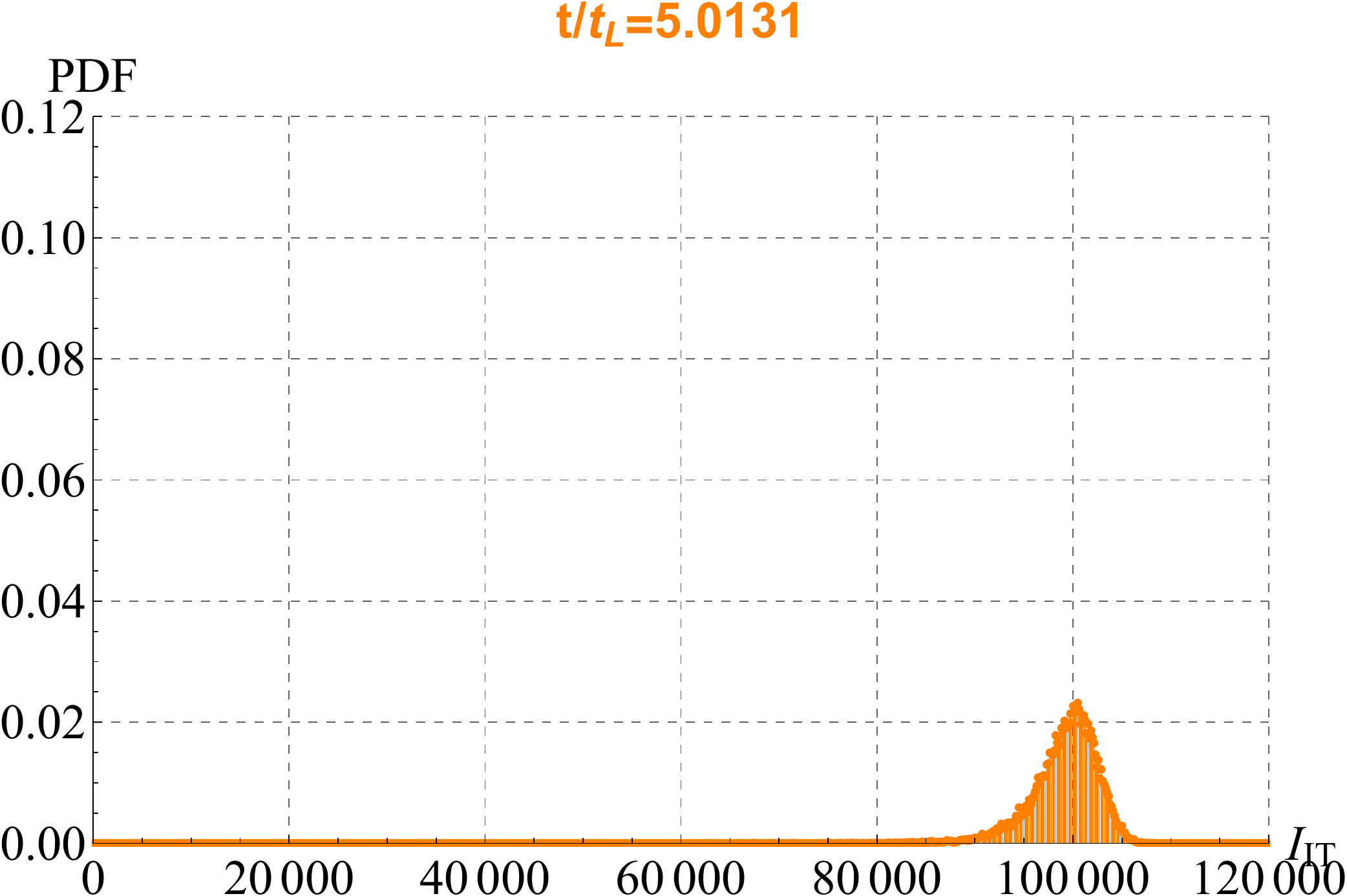}
    \caption{
    \textit{{\bf Probability density function (PDF), solution of Eq.~(\ref{S11}) for Italy, first wave of infection by SARS-CoV-2.} Screenshoot at time  ${\hat t}=5.0131$.} 
    }
    \label{SMF5}
  \end{minipage}
  \hfill
    \begin{minipage}[t]{.45\textwidth}
      \centering
      \includegraphics[width=5cm,height=5cm]{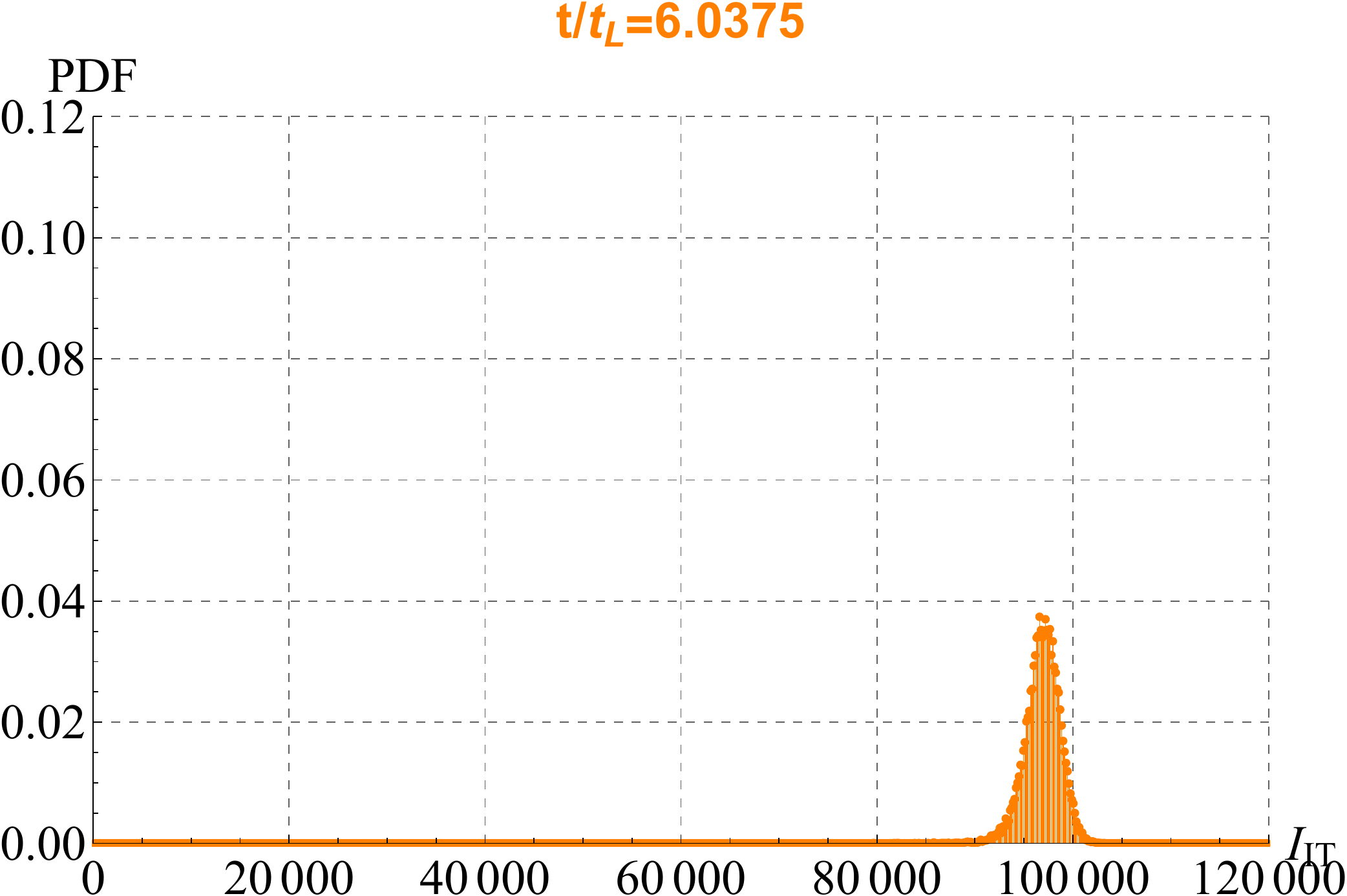}
      \caption{
\textit{{\bf Probability density function (PDF), solution of Eq.~(\ref{S11}).} Screenshoot at time ${\hat t}=6.0375$.} 
      }
      \label{SMF6}
    \end{minipage}
  \hfill
\end{figure*}
\begin{figure*}[htb]
  \hfill
  \begin{minipage}[t]{.45\textwidth}
    \centering
    \includegraphics[width=5cm,height=5cm]{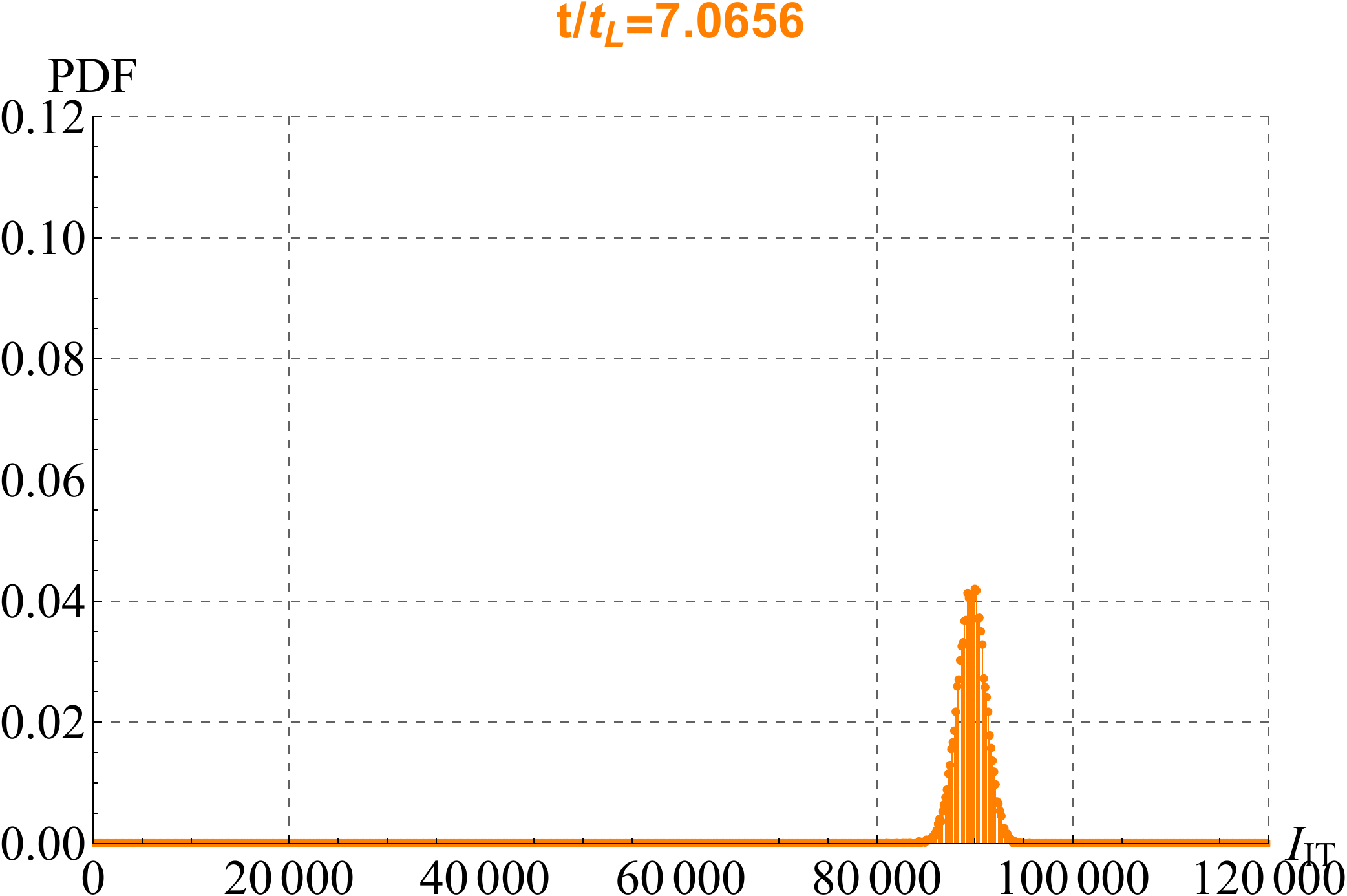}
    \caption{
    \textit{{\bf Probability density function (PDF), solution of Eq.~(\ref{S11}) for Italy, first wave of infection by SARS-CoV-2.} Screenshoot at time  ${\hat t}=7.0656$.} 
    }
    \label{SMF7}
  \end{minipage}
  \hfill
    \begin{minipage}[t]{.45\textwidth}
      \centering
      \includegraphics[width=5cm,height=5cm]{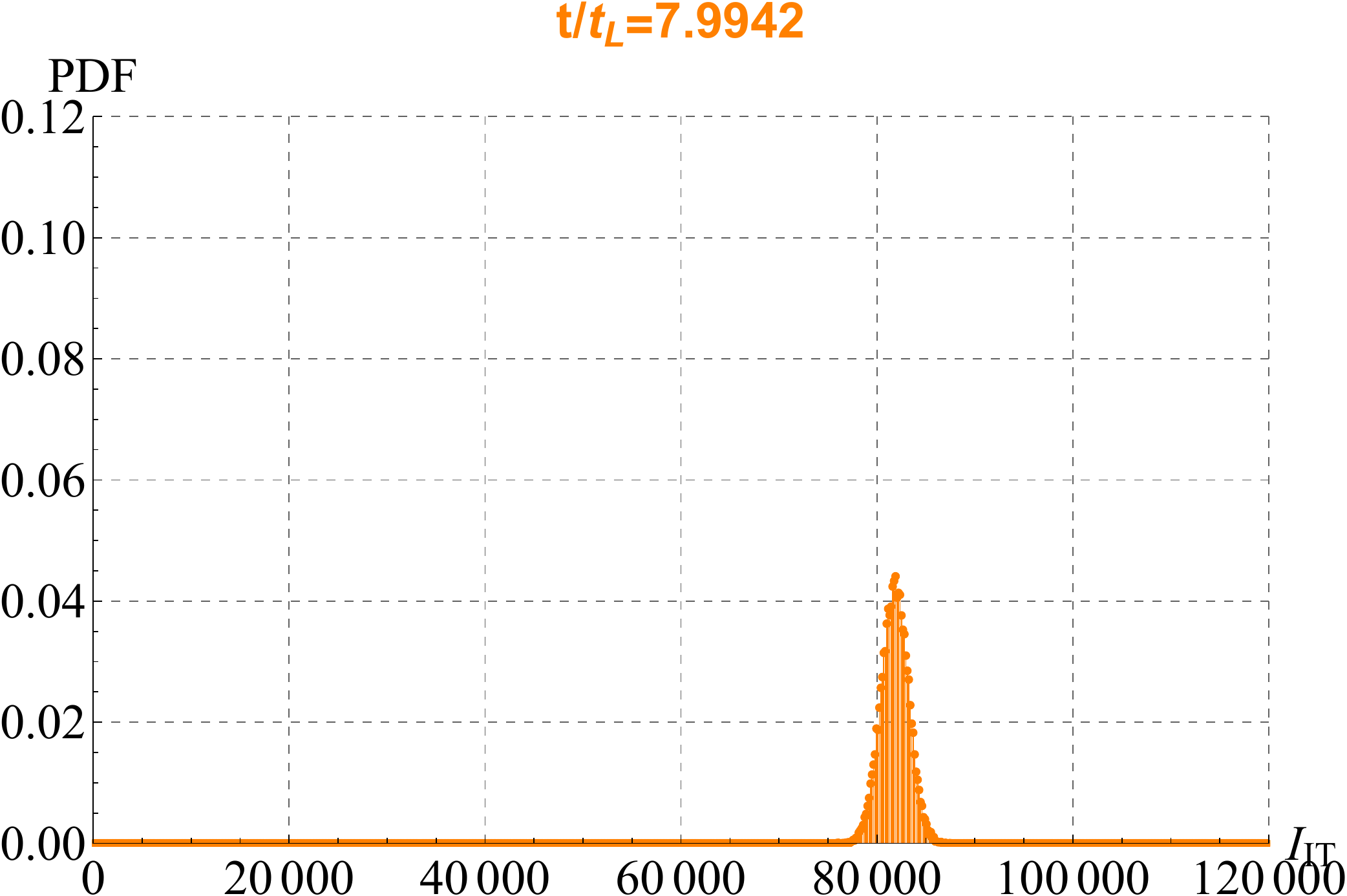}
      \caption{
\textit{{\bf Probability density function (PDF), solution of Eq.~(\ref{S11}).} Screenshoot at time ${\hat t}=7.9942$.} 
      }
      \label{SMF8}
    \end{minipage}
  \hfill
\end{figure*}

\section{\bf The $(SISI_hr_hd_h)_L$-Model}$\label{DM2}$
\noindent As seen, according to the $SIS$-model, after the lockdown measures the number of infectious people starts to increase again. Hospitals and health institutions play a crucial role in hindering the spread of the Coronavirus. In this Section, we propose a model that accounts for people who are only traced back to hospitalised infectious individuals. However, we anticipate that the model for the hospitalised people should be considered as an \textit{purely idealistic}, rather than a \textit{realistic}. model. Finally, we propose the following model\footnote{In refs.~\cite{sonnino2} and \cite{sonnino3} it is proposed to adopt the \textit{Michaelis-Menten's enzyme-substrate reaction model} (the MM reaction) for describing the dynamics of the recovered people previously hospitalised. In our case, the enzyme is compared to the available hospital beds, the substrate to the infected people, and the product to the recovered people, respectively. Of course, the reverse \textit{MM reaction} has no sense in our case and, consequently, the \textit{kinetic constant} is equal to zero.}:
\begin{align}\label{H1}
&S+I \xrightarrow{\sigma} 2I\\
&I\xrightarrow{\gamma+c(t)}S\nonumber\\
&I + b \xrightarrow{k_1} I_h \xrightarrow{k_r} r_h+b\nonumber\\
&\qquad\qquad\ \! I_h \xrightarrow{k_d} d_h+b\nonumber\\
&\qquad\qquad\ \! r_h \xrightarrow{\gamma_1} S\nonumber
\end{align}
\noindent where the hypothesis that an individual acquires immunity, after having contracted the Coronavirus and being recovered, is not adopted. In the scheme~(\ref{H1}), $b$ denotes the number of available \textit{hospital beds}, $I_h$ the number of \textit{infected people blocking an hospital bed}, $r_h$ the number of \textit{recovered people previously hospitalised}, and $d_h$ the number of \textit{people deceased in the hospital}, respectively. In the scheme~(\ref{H1}), we have into account that people, once recovered, are subject to the same existing lockdown measures as any other people.. Of course, we have
\begin{equation}\label{H2}
I_h+b=C_h=const.\qquad {\rm where}\quad{C_h={\rm Total\ hospital's\ capacity}}
\end{equation}
\noindent The dynamic equations for the entire process are then:
\begin{align}\label{H3}
&\frac{d}{d{\hat t}}S=-\sigma\frac{S}{N_{Tot.}} I+\gamma I+\gamma_1 r_h+\beta\left(\frac{{\hat t}^2-1}{\hat t}\right) (I+r_h)\\
&\frac{d}{d{\hat t}}I=\sigma\frac{S}{N_{Tot.}} I-\gamma I-k_1I(C_h-I_h)-\beta\left(\frac{{\hat t}^2-1}{\hat t}\right) I\nonumber\\
&\frac{d}{d{\hat t}}I_h= k_1I(C_h-I_h)-k_r{I_h}-k_d{I_h}\nonumber\\
&\frac{d}{d{\hat t}}r_h=k_r{I_h}-\gamma_1 r_h\nonumber\\
&\frac{d}{d{\hat t}}d_h=k_d {I_h}\nonumber
\end{align}
\noindent where, for simplicity, the \textit{average recovery time delay} and the \textit{average death time delay} have been neglected \cite{sonnino2}. In this case $I$ stands for the \textit{infectious individuals not hospitalised}. From system~(\ref{H3}) we get the following conservation law
\begin{equation}\label{H4}
S+I+I_h+r_h+d_h=N_{Tot.}=const.
\end{equation}
\subsection{\bf The Deterministic $(SISI_h)_L$-Model}\label{ASM}
\noindent To simplify as much as possible the set of O.D.E.s~(\ref{H3}), we adopt several hypotheses that will not compromise the validity of our model. First, we assume that that $S+I+I_h\simeq N_{Tot.}=const.$. Secondly, let $\gamma\simeq\gamma_1$. Finally, we take into account the current Belgian hospital-protocol: "\textit{Only the seriously sick people are hospitalised, the remaining infectious individuals have to be sent home and they must be subject to quarantine measures}". Hence, $I\gg I_h $ and the total number of recovered people, $R$, is much larger than the total number of the recovered people, previously hospitalised (i.e. $R\gg r_h$). Under these assumptions, the model simplifies to
\begin{align}\label{AH1}
&S+I \xrightarrow{\sigma} 2I\\
&I\xrightarrow{\gamma+c(t)}S\nonumber\\
&I + b \xrightarrow{k_1} I_h
\end{align}
\noindent with $I_h+b=C_h$. Hence, under these assumptions, after hospitalisation, individuals will be removed from the disease, either due to immunisation (e.g. due to vaccination or special health care received) or due to death. The governing O.D.E.s, associated to the model~(\ref{AH1}), read
\begin{align}\label{AH2}
&\frac{d}{d{\hat t}}S\simeq-\sigma\frac{S}{N_{Tot.}} I+\gamma I+\beta\left(\frac{{\hat t}^2-1}{\hat t}\right) I\\
&\frac{d}{d{\hat t}}I=\sigma\frac{S}{N_{Tot.}} I-\gamma I-k_1I(C_h-I_h)-\beta\left(\frac{{\hat t}^2-1}{\hat t}\right) I\nonumber\\
&\frac{d}{d{\hat t}}I_h\simeq k_1I(C_h-I_h)\nonumber
\end{align}
\noindent The recovered people $r_h$ and the deceased people $d_h$ may be obtained by solving, respectively, the following O.D.E.s
\begin{equation}\label{AH3}
\frac{d}{d{\hat t}}r_h\simeq k_r{I_h}\qquad ;\qquad \frac{d}{d{\hat t}}d_h=k_d {I_h}
\end{equation}
\noindent From system~(\ref{AH2}), we obtain 
\begin{align}\label{AH4}
&\frac{d}{d{\hat t}}I=\alpha_1I\left(1-\frac{I}{K_1}\right)-\rho I I_h-\beta\left(\frac{{\hat t}^2-1}{\hat t}\right) I\\
&\frac{d}{d{\hat t}}I_h= k_1C_h I-k_1I I_h\qquad{\rm with}\nonumber\\
&\alpha_1=\sigma\!\left(\!1-\frac{\gamma}{\sigma}-\frac{k_1}{\sigma}C_h\!\right)\ ;\ K_1=N_{Tot.}\left(\!1-\frac{\gamma}{\sigma}-\frac{k_1}{\sigma}C_h\!\right)\ ;\ \rho=\frac{\sigma}{N_{Tot.}}\!-\!k_1\nonumber
\end{align}
\noindent In absence of the lockdown measures ($\beta=0$), we have the following scenarios: $\forall I_0>0$
\begin{align}\label{AH5}
&{\bf i)}\ \ {\rm if}\quad C_h<C_{hCrit.}\equiv\frac{\sigma}{k_1}\left(1-\frac{\gamma}{\sigma}\right)\  {\rm the\ equilibrium\ with}\ I=K_1\ {\rm is\ stable}\\
&{\bf ii)}\ \ \ {\rm if}\quad C_h>C_{hCrit.}\equiv\frac{\sigma}{k_1}\left(1-\frac{\gamma}{\sigma}\right)\  {\rm the\ equilibrium\ with}\ I=0\ {\rm is\ stable}
\nonumber
\end{align}
\noindent In words: 

\noindent $\bullet$ for the case ${\bf i}$), there will be a proper epidemic outbreak with an increase of the number of the infectious people;

\noindent $\bullet$ for the case ${\bf ii}$), independently of the initial size of the susceptible population, the disease can never cause a proper epidemic outbreak.

\noindent This result highlights the crucial role of the Hospitals and the Health Care Institutes: 

\noindent \textit{If the threshold of the hospital capacities exceeds a minimum limit, the spread of the Coronavirus tends to decrease over time, and the stable solution corresponds to zero infectious individuals}.

\noindent Fig.~\ref{H1}. and \ref{H0}. illustrate the situation. Notice that ,with the values of parameters reported in the figure captions, $C_{hCrit.}=18434$.
\begin{figure*}[htb]
  \hfill
  \begin{minipage}[t]{.45\textwidth}
    \centering
    \includegraphics[width=4cm,height=3cm]{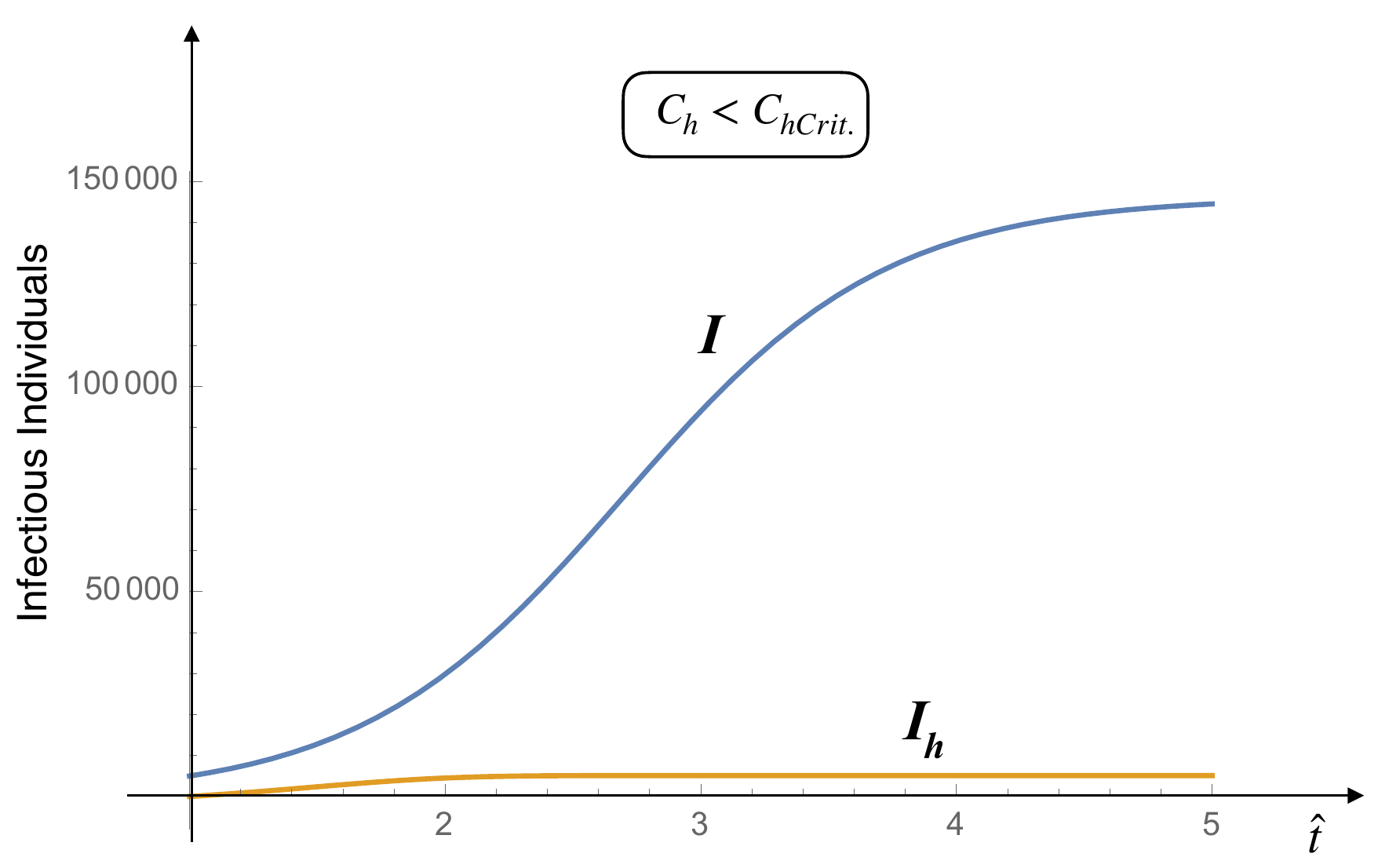}
    \caption{
    \textit{{\bf Infectious people and Infectious hospitalised people vs time - $C_h>C_{HCrit.}$}. In this case, there is an epidemic outbreak with an increase of the number of the infectious people. The values of the parameters are: $\beta=0$, $k_1=0.00001$, $\rho=0.0001$, $K_1=150000$, $\gamma=0.0001$, $\sigma=2.7651$, and $C_h=5100$ }.
    }
    \label{H1}
  \end{minipage}
  \hfill
    \begin{minipage}[t]{.45\textwidth}
      \centering
      \includegraphics[width=4cm,height=3cm]{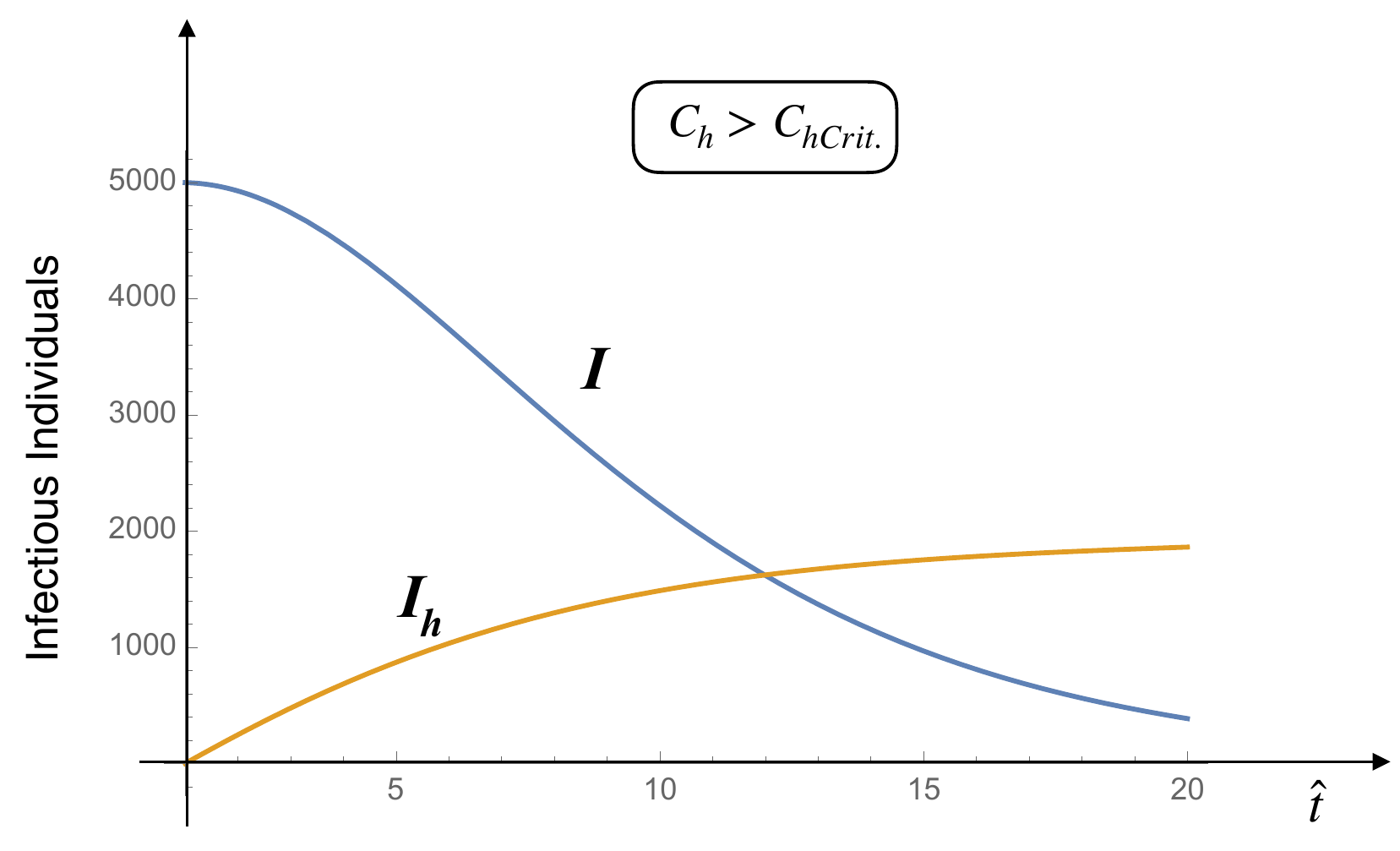}
      \caption{
\textit{{\bf Infectious people and Infectious hospitalised people vs time - $C_h>C_{HCrit.}$}. Independently of the initial size of the susceptible population, the disease can never cause a proper epidemic outbreak. The values of the parameters are: $\beta=0$, $k_1=0.00001$, $\rho=0.0001$, $K_1=150000$, $\gamma=0.0001$, $\sigma=2.7651$, and $C_h=18436$}.
      }
      \label{H0}
    \end{minipage}
  \hfill
\end{figure*}
\section {\bf The Stochastic $(SISI_h)_L$ Model}\label{SMS}

\noindent If the dynamics is subject to a white noise, the related stochastic equations read
\begin{align}\label{SM1}
&\frac{d}{d{\hat t}}I=\alpha_1I\left(1-\frac{I}{K_1}\right)-\rho I I_h-\beta\left(\frac{{\hat t}^2-1}{\hat t}\right) I+\xi_1({\hat t})\\
&\frac{d}{d{\hat t}}I_h= k_1C_h I-k_1I I_h+\xi_2({\hat t})\nonumber
\end{align}
\noindent where
\begin{align}\label{SM2}
&<\xi_i(t)>=0\qquad (i=1,2)\\
&<\xi_i(t)\xi_j(t')>=\eta_{ij}\delta_{ij}\delta(t-t')\qquad{\rm with}\quad \eta_{12}=\eta_{21}\nonumber
\end{align}
\noindent with $\delta_{ij}$ denoting Kronecker's delta. The statistical properties of this processes, i.e. $<\delta I({\hat t})\delta I({\hat t})>$, $<\delta I({\hat t})\delta I_h({\hat t})>$ and $<\delta I_h({\hat t})\delta I_h({\hat t})>$, may be obtained, firstly, by determining the reference state. This state satisfies the following O.D.E.s
\begin{align}\label{SM3}
&\frac{d}{d{\hat t}}I_{RS}=\alpha_1I_{RS}\left(1-\frac{I_{RS}}{K_1}\right)-\rho I_{RS} I_{hRS}-\beta\left(\frac{{\hat t}^2-1}{\hat t}\right) I_{RS}\\
&\frac{d}{d{\hat t}}I_{hRS}= k_1C_h I_{RS}-k_1I_{RS} I_{hRS}\nonumber
\end{align}
\noindent Let us choose, for example, $I_{hRS}=C_h$, then 
\begin{align}\label{SM4}
&I_{RS}({\hat t})=\frac{I_{0RS}\exp((1-\alpha_2/\beta)^2/\sigma){\hat t}^\beta\exp(-({\hat t}-\alpha_2/\beta)^2/\sigma)}{1+(I_{0RS}\alpha_2/K_2)\exp((1-\alpha_2/\beta)^2/\sigma)\int_1^{\hat{t}}x^\beta\exp(-(x-\alpha_2/\beta)^2/\sigma)dx}\nonumber\\
&\alpha_2\equiv\sigma\left(1-\frac{\gamma}{\sigma}-\frac{C_h}{N_{Tot.}}\right)\quad ;\quad K_2=N_{Tot.}\left(1-\frac{\gamma}{\sigma}-\frac{C_h}{N_{Tot.}}\right)
\end{align}
\noindent Successively, we have to find the solution of the Fokker-Planck equation, $p(I,I_h,{\hat t})$, associated to the system~(\ref{SM1}), which in this case reads
\begin{equation}\label{SM5}
\frac{d}{d{\hat t}}p=-\frac{\partial}{\partial I}(\mu_1p)-\frac{\partial}{\partial I_h}(\mu_2p)+D_{11}\frac{\partial}{\partial I^2}p+D_{12}\frac{\partial}{\partial I\partial I_h}p+D_{22}\frac{\partial}{\partial I_h^2}p
\end{equation}
\noindent with
\begin{align}\label{SM6}
&\mu_1=\alpha_1I\left(1-\frac{I}{K_1}\right)-\rho I I_h-\beta\left(\frac{{\hat t}^2-1}{\hat t}\right) I\quad ;\quad \mu_2= k_1C_h I-k_1I I_h\\
&D_{11}=\frac{1}{2}(\eta_{11}+\eta_{12})\ ;\ D_{12}=\sqrt{\eta_{12}}(\sqrt{\eta_{11}}+\sqrt{\eta_{22}})\ ;\ D_{22}=\frac{1}{2}(\eta_{22}+\eta_{12})\nonumber
\end{align}
\noindent Finally, with the expression of $p(I,I_h,t)$ we may compute the correlation functions $<\delta I({\hat t})\delta I({\hat t})>$, $<\delta I({\hat t})\delta I_h({\hat t})>$ and $<\delta I_h({\hat t})\delta I_h({\hat t})>$ with
\begin{equation}\label{SM7}
\delta I=I-I_{RS}\qquad ;\qquad \delta I_h=I_h-I_{hRS}
\end{equation}
\noindent However, this is a quite long and tedious procedure. We may circumvent the obstacle by recalling that, as we did in Section~(\ref{SM}), it is possible to obtain such correlation functions in a \textit{direct way}, by using the theorem~(\ref{S5}-\ref{S6}). For this, we have to determine the dynamic O.D.E.s for the fluctuations $\delta I({\hat t})$ and $\delta I_h({\hat t})$ around the \textit{reference state} ($I_{RS}, I_{hRS}$). We get
\begin{align}\label{SM8}
&\frac{d}{d{\hat t}}\delta I=\left(\alpha_1-\rho C_h-2\frac{\alpha_1}{K_1}I_{RS}-\beta\left(\frac{{\hat t}^2-1}{\hat t}\right)\right)\delta I-\rho I_{RS}\delta I_h+\zeta_1({\hat t})\nonumber\\
&\frac{d}{d{\hat t}}\delta I_h=-k_1I_{RS}(t)\delta I_h+\zeta_2({\hat t})
\end{align} 
\noindent The correlation functions $<\delta I({\hat t}_0)\delta I({\hat t})>$ and  $<\delta I_h({\hat t}_0)\delta I_h({\hat t})>$ are easily obtained 
 \begin{align}\label{SM9}
 &<\!\delta I({\hat t}_0)\delta I({\hat t})\!>\!=\!\left(\frac{{\hat t}}{{\hat t}_0}\right)^\beta\exp(H({\hat t}))\exp\left(-2\frac{\alpha_1}{K_1}G({\hat t}))\right)\!\!\Bigg(\!\!<\!\delta I({\hat t}_0)^2\!>\\
 &-\rho<\delta I({\hat t}_0)\delta I_h({\hat t}_0)>\int_{{\hat t}_0}^{\hat t}\left(\frac{{\hat t}_0}{s}\right)^\beta\! I_{RS}(s)\exp(-H(s))\exp\left((2\frac{\alpha_1}{K_1}-k_1)G(s)\right)\!ds\Bigg)\nonumber\\
&<\delta I_h({\hat t}_0)\delta I_h({\hat t})>=<\delta I_h({\hat t}_0)^2>\exp(-k_1 G({\hat t}))\qquad{\rm with}\nonumber\\
&G({\hat t})\equiv\int_{{\hat t}_0}^{\hat t}I_{RS}(s)ds\quad;\quad H({\hat t})\equiv (\alpha_1-\rho C_h)({\hat t}-{\hat t}_0)-1/2\beta({\hat t}^2-{\hat t}_0^2)
\nonumber
\end{align} 
 \noindent In addition, by using again Eqs.~(\ref{SM8}), we obtain the O.D.E.s for the correlation functions for $<(\delta I({\hat t}))^2>$, $<(\delta I_h({\hat t}))^2>$, and $<\delta I({\hat t})\delta I_h({\hat t})>$
 \begin{align}\label{SM10}
&\frac{d}{d{\hat t}}<(\delta I({\hat t}))^2>=2\left(\alpha_1-\rho C_h-2\frac{\alpha_1}{K_1}I_{RS}-\beta\left(\frac{{\hat t}^2-1}{\hat t}\right)\right)<(\delta I({\hat t}))^2>\\
&\qquad\qquad\qquad\qquad\ \ -2\rho I_{RS}<\delta I({\hat t})\delta I_h({\hat t})>+\eta_{11}({\hat t})\nonumber\\
&\frac{d}{d{\hat t}}<(\delta I_h({\hat t}))^2>=-2k_1I_{RS}<(\delta I_h({\hat t}))^2>+\eta_{22}\nonumber\\
&\frac{d}{d{\hat t}}\!<\!\delta I({\hat t})\delta I_h({\hat t})\!>=\!\left(\!\alpha_1\!-\!\rho C_h\!-\left(\!2\frac{\alpha_1}{K_1}+k_1\!\right)I_{RS}\!-\!\beta\left(\!\frac{{\hat t}^2\!-\!1}{\hat t}\right)\!\right)\!\!<\!\delta I({\hat t})\delta I_h({\hat t})\!> \nonumber\\
&\qquad\qquad\qquad\qquad\ \  -\rho I_{RS}<(\delta I_h({\hat t}))^2>+\eta_{12}({\hat t})\nonumber
\end{align}
\noindent where Eqs~(\ref{S6}) have been taken into account. The solutions of system~(\ref{SM10}) read
\begin{align}\label{SM11}
&<\!(\delta I_h({\hat t}))^2\!>\!=\!-\exp(-2k_1G({\hat t}))\!\left(\!<\!(\delta I_h({\hat t}_0))^2\!>\!+\eta_{22}\!\int_{{\hat t}_0}^{\hat t}\!\! \exp(2k1G(s)\!ds\right)\\
&<\!\delta I({\hat t})\delta I_h({\hat t})\!>=\!\left(\frac{{\hat t}}{{\hat t}_0}\right)^\beta\!\!\!\exp(H({\hat t}))\!\exp\left(-(2\alpha_1/K_1+k_1)\right)\!G({\hat t}))\!\Bigg(\!\!\!<\delta I({\hat t}_0)\delta I_h({\hat t}_0)\!>\nonumber\\
&-\int_{{\hat t}_0}^{\hat t}\!\!\left(\frac{{\hat t}_0}{s}\right)^\beta\!\!\!\!\exp(\!-\!H(s))\exp\left((2\alpha_1/K1\!+\!k_1)G(s)\right)\!\!\left(\eta_{12}\!+\!\rho I_{RS}(s)<\!(\delta I_h(s))^2\!>\right)\!ds\Bigg)\nonumber\\
&<(\delta I({\hat t}))^2>=\left(\frac{{\hat t}}{{\hat t}_0}\right)^{2\beta}\exp(2H({\hat t}))\exp\left(-4\frac{\alpha_1}{K_1}\right)G({\hat t}))\Bigg(<(\delta I({\hat t}_0))^2>\nonumber\\
&-\int_{{\hat t}_0}^{\hat t}\!\!\left(\frac{{\hat t}_0}{s}\right)^{2\beta}\!\!\!\!\!\!\exp(\!-\!2H(s))\!\exp\left((4\alpha_1/K1\!)G(s)\right)\!\left(\eta_{11}\!\!-2\!\rho I_{RS}(s)<\!\delta I(s)\delta I_h(s)\!>\right)\!ds\Bigg)\nonumber\
\nonumber
\end{align}
\noindent The value of $C_h$ (the total Italian hospitals' capacity) may be obtained by making reference to the data published in \cite{hospitalsIT}. More specifically, in 2017, when there were $518$ public hospitals and $482$ accredited private ones, in Italy there were $151646$ beds for ordinary hospitalisation in public hospitals ($2.5$ per $1000$ inhabitants) and $40458$ in private ones ($0.7$ per $1000$ inhabitants), for a total of over $192$ thousand beds ($3.2$ per $1000$ inhabitants).

\noindent The number of public and private beds destined for intensive care was 5.090 (a number very close to the $5100$ cited by the newspapers these days), about $8.42$ per $100000$ inhabitants \cite{hospitalsIT}. 

\noindent Fig.~\ref{CFIhIh}. shows the correlation function $<\delta I_h(t)\delta I_h(t)>$, solution of the second O.D.E. of system~(\ref{SM10}). As se can see, this is a typical correlation function that random fluctuations \textit{at equilibrium} look like. This is not surprising as our reference state correspond to the maximum capacity of the hospitals (i.e. $I_ {hRS}=C_h$) so, fluctuations at equilibrium are the only possible ones. The three correlation functions $<\delta I(t)\delta I(t)>$, $<\delta I_h(t)\delta I_h(t)>$, and $<\delta I(t)\delta I_h(t)>$, solution of the system~(\ref{SM10}), are reported in Fig.~\ref{CFs}.
\begin{figure}[hbt!]
\hskip 2truecm
\includegraphics[width=8cm, height=6cm]{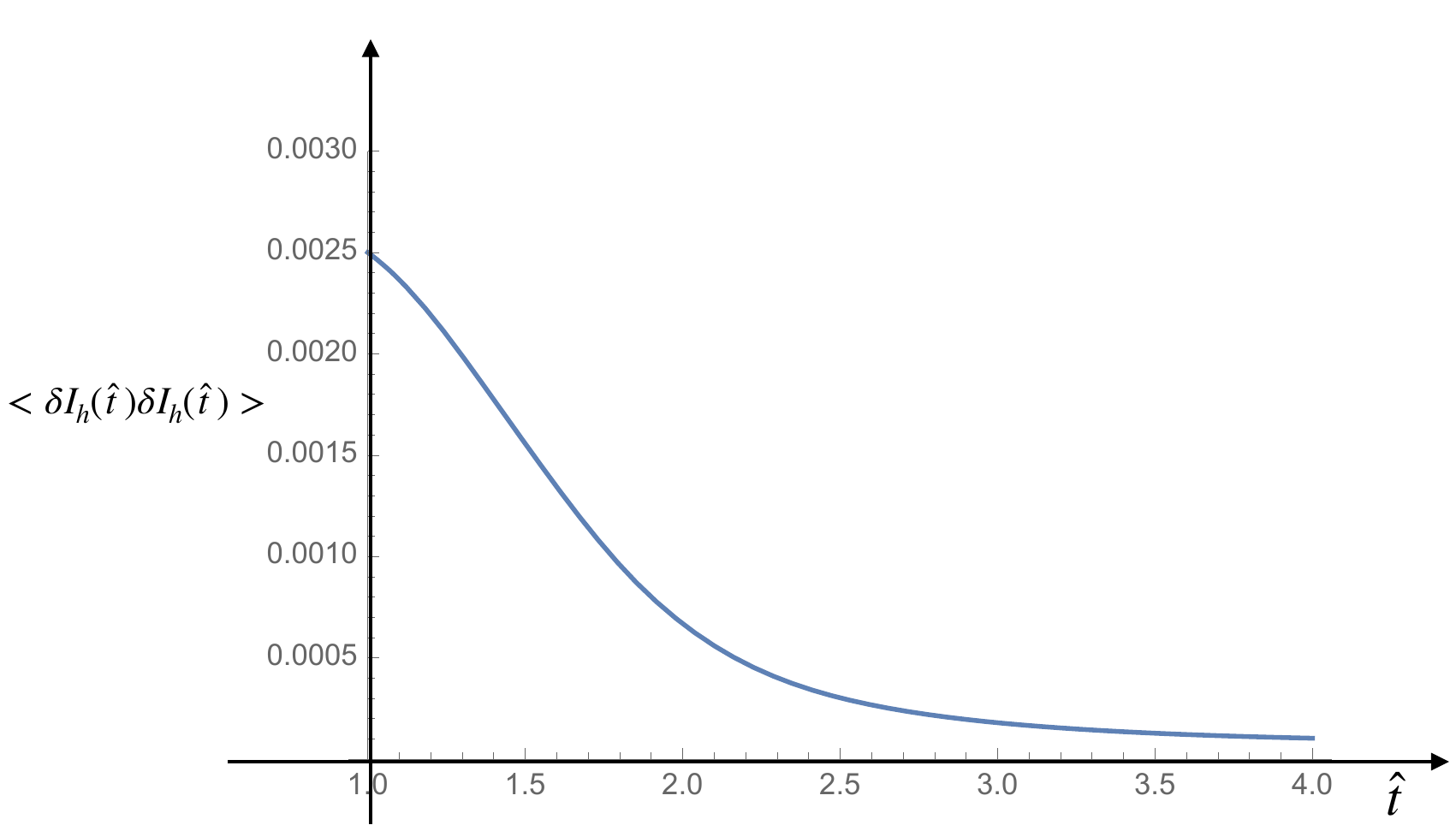}
\caption{
\textit{{\bf Correlation function $<(\delta I_h(t))^2>$ for Italy, first wave of infection by SARS-CoV-2.} This correlation function is the solution of second equation of the system~(\ref{SM10}). Having chosen our reference state $I_{hRS}=C_h$,  we get a typical correlation function for random fluctuations \textit{at equilibrium}.} 
}
\label{CFIhIh}
\end{figure}
\begin{figure}[hbt!]
\hskip 2truecm
\includegraphics[width=8cm, height=6cm]{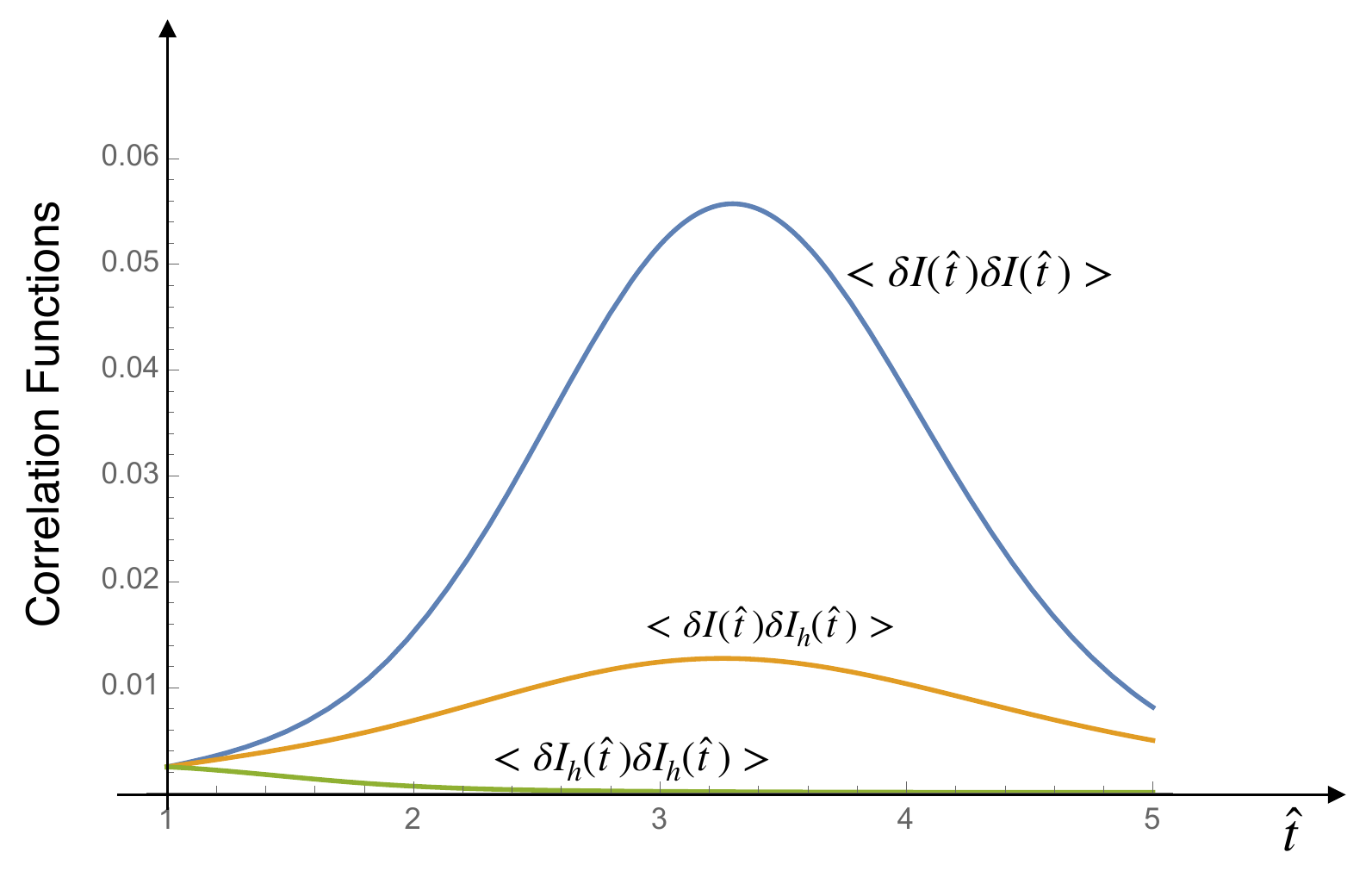}
\caption{
\textit{{\bf Correlation functions} $<\delta I(t)\delta I(t)>$, $<\delta I_h(t)\delta I_h(t)>$, and $<\delta I(t)\delta I_h(t)>$. These three correlation functions are obtained by solving numerically system~(\ref{SM10}).} 
}
\label{CFs}
\end{figure}
\section {\bf Comparison between the Theoretical Predictions and Real Data for USA and France}\label{comp}

\noindent To perform the comparison with real data, it is convenient to rewrite Eq.~(\ref {S1}) in dimensional form
\begin{equation}\label{comp1}
\frac{dI}{d t}={\tilde \alpha} I\left(1-\frac{I}{K}\right)-\beta\left(\frac{t^2-t_L^2}{t_L^2t}\right)I+\ \sqrt{\tilde\eta}\ {\tilde \xi}(t)
\end{equation}
\noindent where ${\tilde\alpha}\equiv{\alpha}/t_L$ and the white noise $\tilde\xi(t)$ satisfies the relations
\begin{align}\label{comp2}
&<{\tilde\xi}(t)>=0\\
&<{\tilde\xi}(t){\tilde\xi}(t')>=\delta(t-t')\nonumber
\end{align}
\noindent The order of magnitude of the intensity of the noise $\tilde\eta$ may be derived by using Eq.~(\ref{A11}) reported in Appendix 1:
\begin{equation}\label{comp3}
{\tilde\eta}\sim K^{-1/2}
\end{equation}
\noindent Appendix 2 reports the tables of infectious people by SARS-CoV-2 for USA and France according to refs~\cite{usa}, \cite{usasituation3} and \cite{france}, \cite{france1}, respectively. As the data show, France is currently subject to the second wave of Coronavirus while the USA data may induce to think that they are in full second (or third) wave. However, looking at the behaviour of infectious curve we may also argue that, in agreement with ref.~\cite{usasituation2}, \textit{USA as a whole is not in a second (or third) wave because the first wave never really stopped. The virus is simply spreading into new populations or resurgent in places that let down their guard too soon}. So, we are interested in analysing both of these scenarios. 

\noindent Fig.~\ref{USAfw}. and Fig.~\ref{USA3w}. show the comparison between the theoretical predictions (blue line) and real data (black dots) for USA. Fig.~\ref{USAfw}. refers to the hypothesis that US is still in the first wave of Coronavirus, whereas Fig.~\ref{USA3w}. has been obtained by assuming that US population is currently subject to the third wave of Coronavirus, respectively. Fig.~\ref{Francefw}. and Fig.~\ref{Francesw}. show the comparison between the theoretical predictions (blue line) and real data (black dots) for France. The values of the parameters for both cases, USA and France, are reported in the figure captions.
\begin{figure*}[htb]
  \hfill
  \begin{minipage}[t]{.45\textwidth}
    \centering
    \includegraphics[width=5cm,height=5cm]{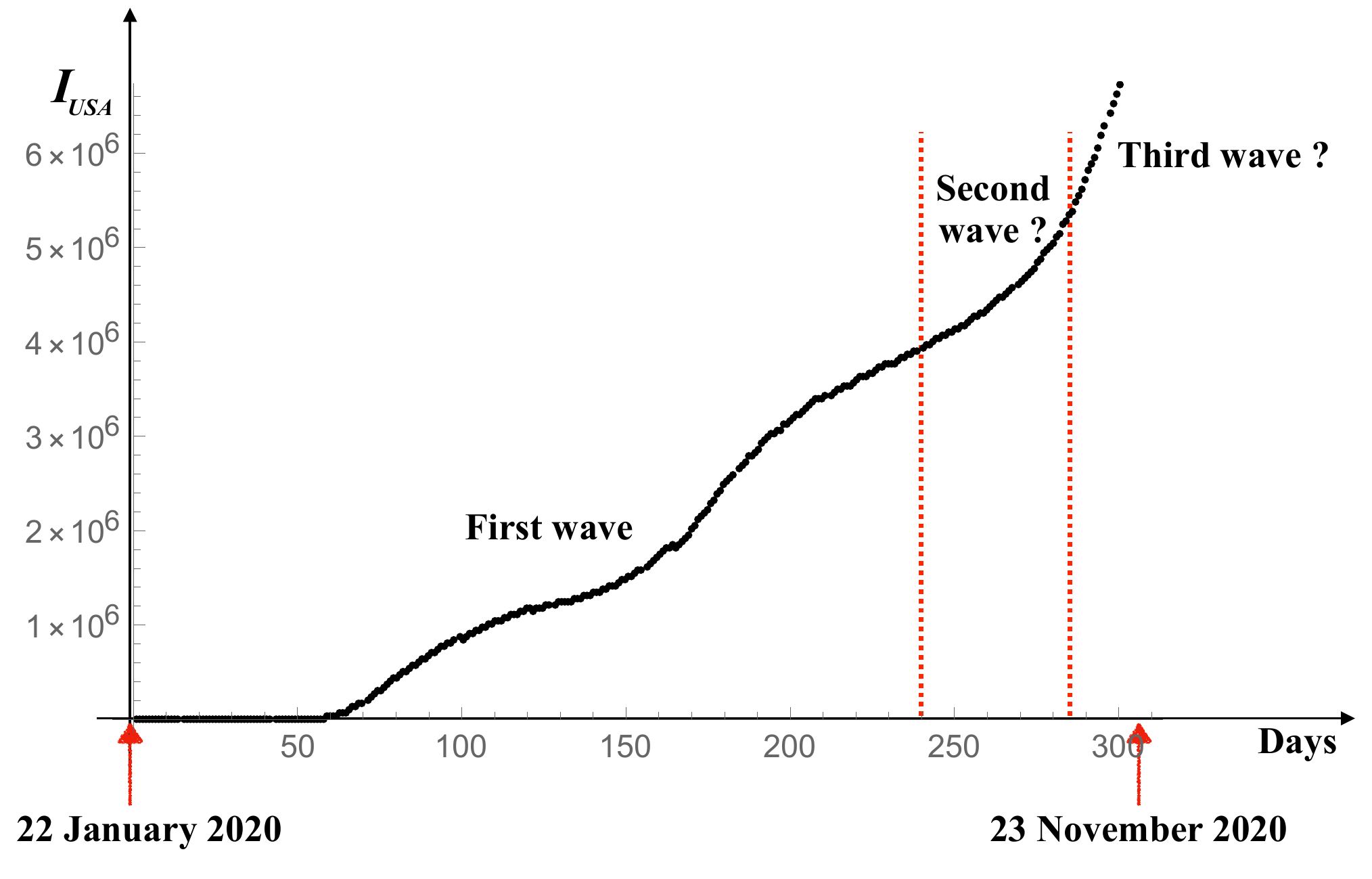}
    \caption{
    \textit{{\bf Number of Infectious individuals by SARS-CoV-2 in USA.} }
    }
    \label{USA_GData}
  \end{minipage}
  \hfill
    \begin{minipage}[t]{.45\textwidth}
      \centering
      \includegraphics[width=5cm,height=5cm]{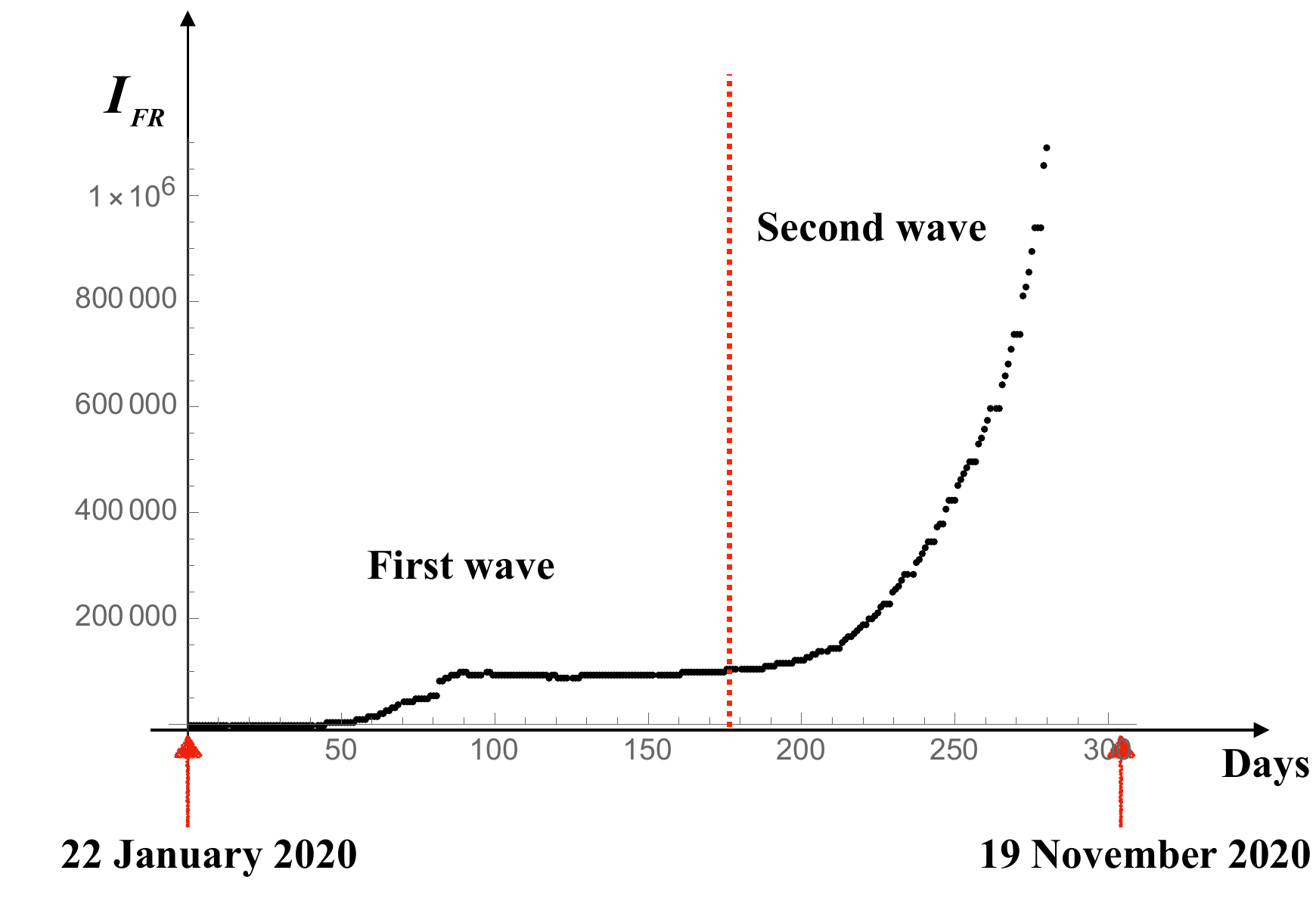}
      \caption{
\textit{{\bf Number of Infectious individuals by SARS-CoV-2 in France.} }
      }
      \label{France_GData}
    \end{minipage}
  \hfill
\end{figure*}
\begin{figure*}[htb]
  \hfill
  \begin{minipage}[t]{.45\textwidth}
    \centering
    \includegraphics[width=5cm,height=5cm]{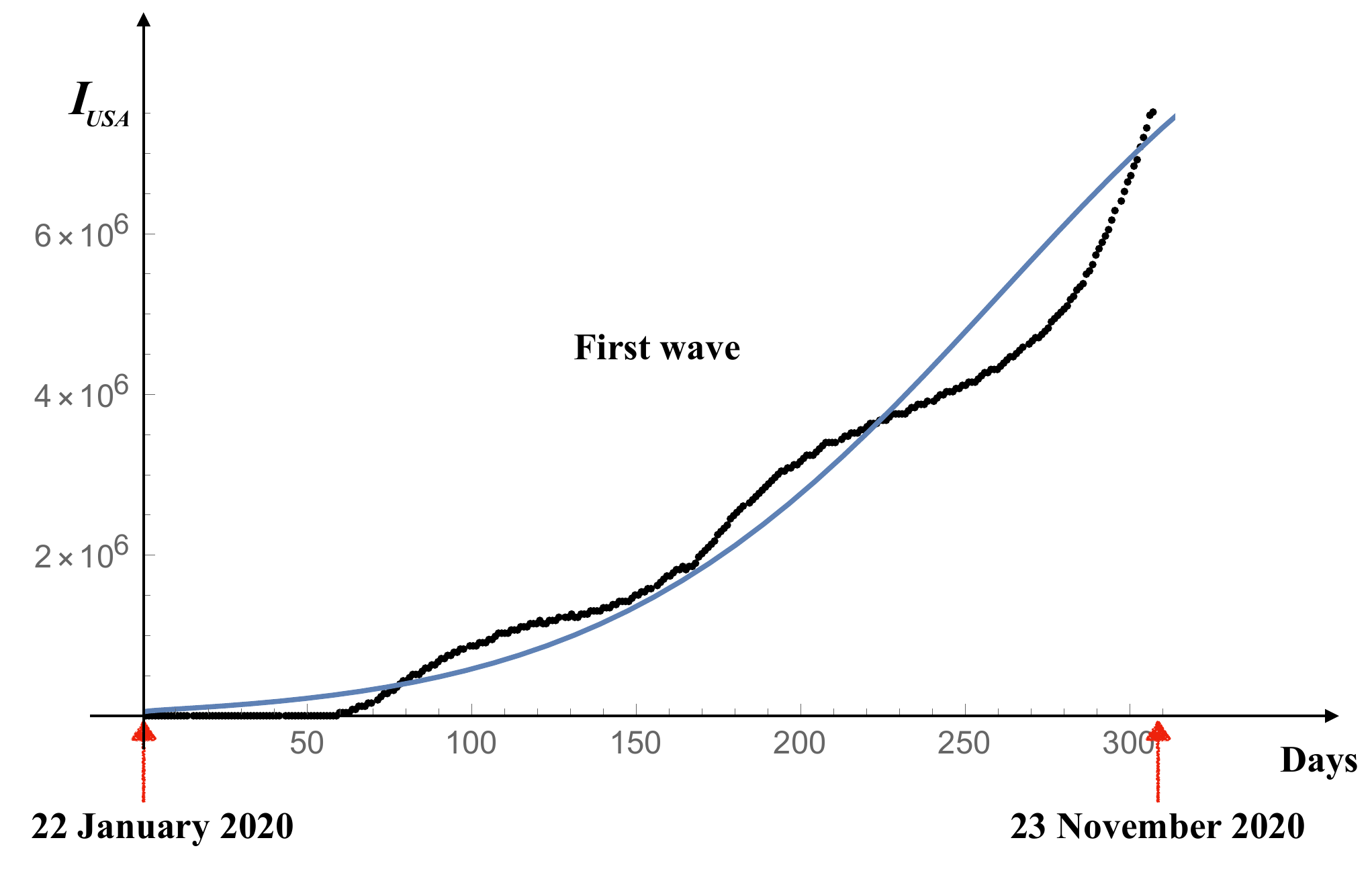}
    \caption{
    \textit{Comparison between the theoretical predictions (blue line) and real data (black dots) for USA by assuming that US is still in the first wave of Coronavirus. The values of the parameters are; ${\tilde\alpha}=0.021\ {\rm day}^{-1}$, $\beta=0.095\ {\rm day}^{-2}$, $K=25000000$, and $t_L=54\ {\rm days}$, respectively.}
    }
    \label{USAfw}
  \end{minipage}
  \hfill
    \begin{minipage}[t]{.45\textwidth}
      \centering
      \includegraphics[width=5cm,height=5cm]{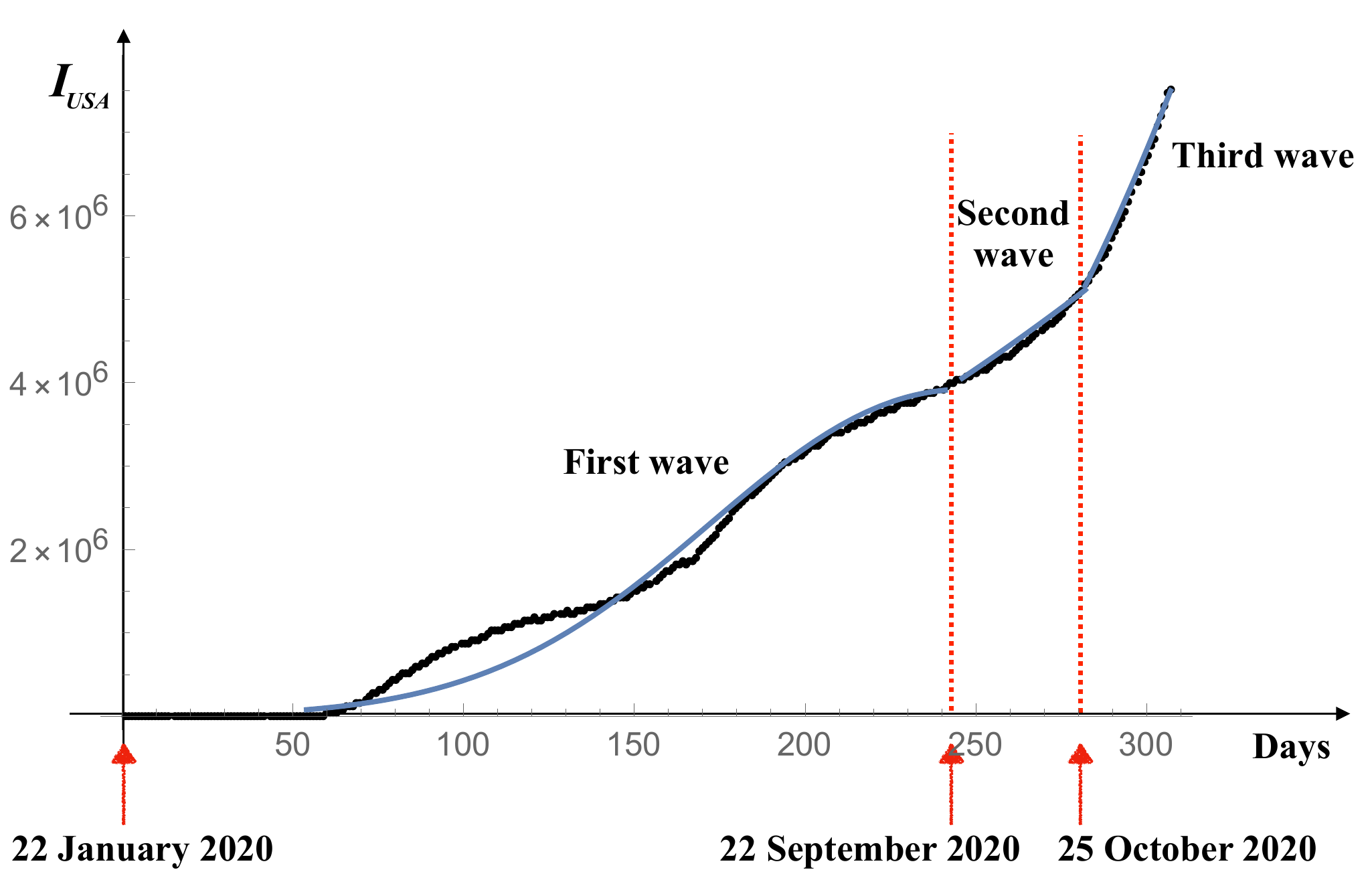}
      \caption{
\textit{Comparison between the theoretical predictions (blue lines) and real data (black dots) for USA by assuming that US is in the third wave of Coronavirus. The three series of values of the parameters, corresponding to the three waves, are, respectively: (${\tilde\alpha}=0.043\ {\rm day}^{-1}$, $\beta=0.45\ {\rm day}^{-2}$) ; (${\tilde\alpha}=0.021\ {\rm day}^{-1}$, $\beta=0.008\ {\rm day}^{-2}$), and (${\tilde\alpha}=0.01\ {\rm day}^{-1}$, $\beta=0.02\ {\rm day}^{-2}$). $K=25000000$ and $t_L=54\ {\rm days}$.}
      }
      \label{USA3w}
    \end{minipage}
  \hfill
\end{figure*}
\begin{figure*}[htb]
  \hfill
  \begin{minipage}[t]{.45\textwidth}
    \centering
    \includegraphics[width=7cm,height=5cm]{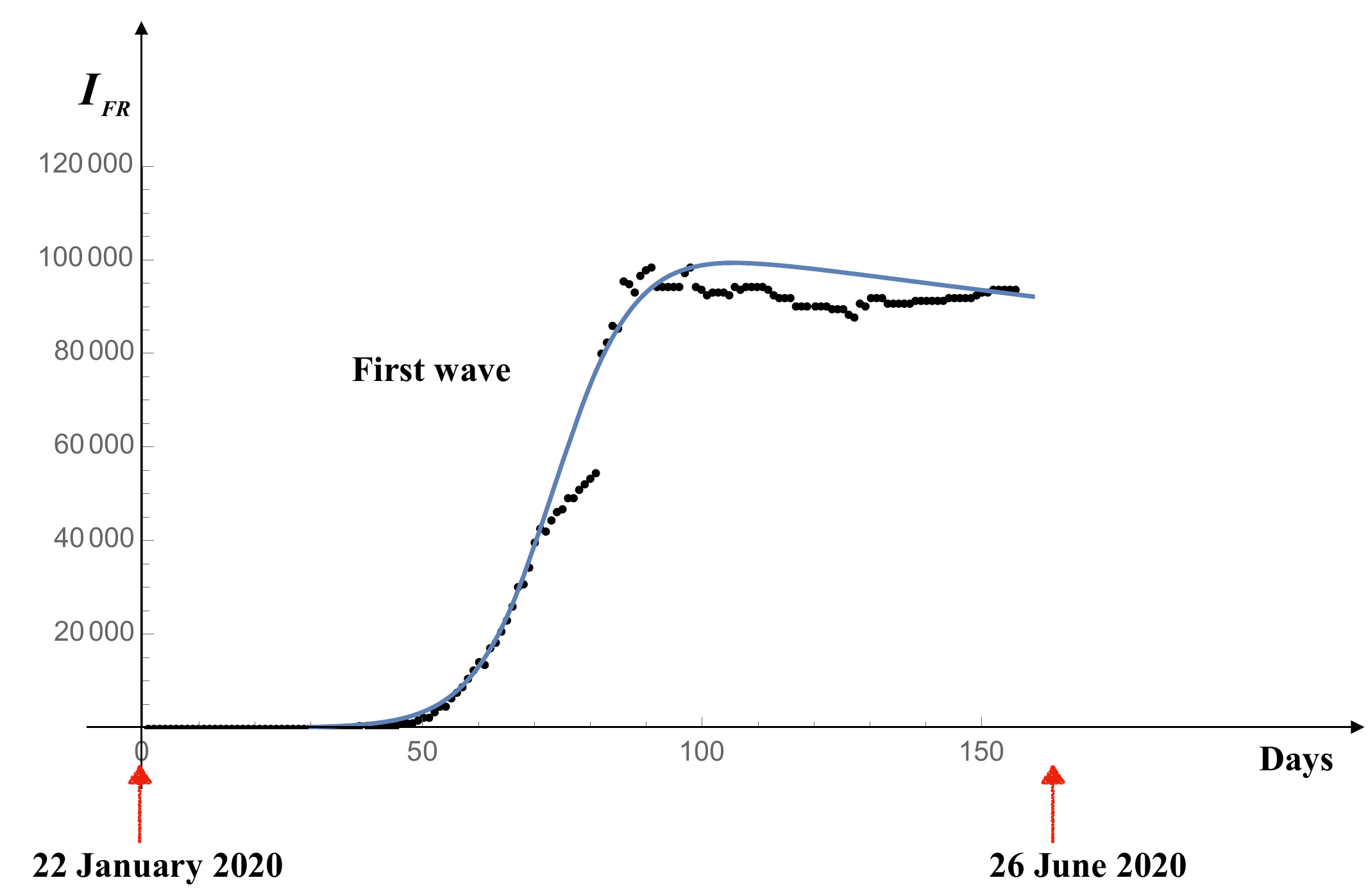}
    \caption{
    \textit{Comparison between the theoretical predictions (blue line) and real data (black dots) for France during the first wave of Coronavirus. The values of the parameters are; ${\tilde\alpha}=0.145\ {\rm day}^{-1}$, $\beta=0.5\ {\rm day}^{-2}$, $K=110000$, and $t_L=53\ {\rm days}$, respectively.}
    }
    \label{Francefw}
  \end{minipage}
  \hfill
    \begin{minipage}[t]{.45\textwidth}
      \centering
      \includegraphics[width=5cm,height=5cm]{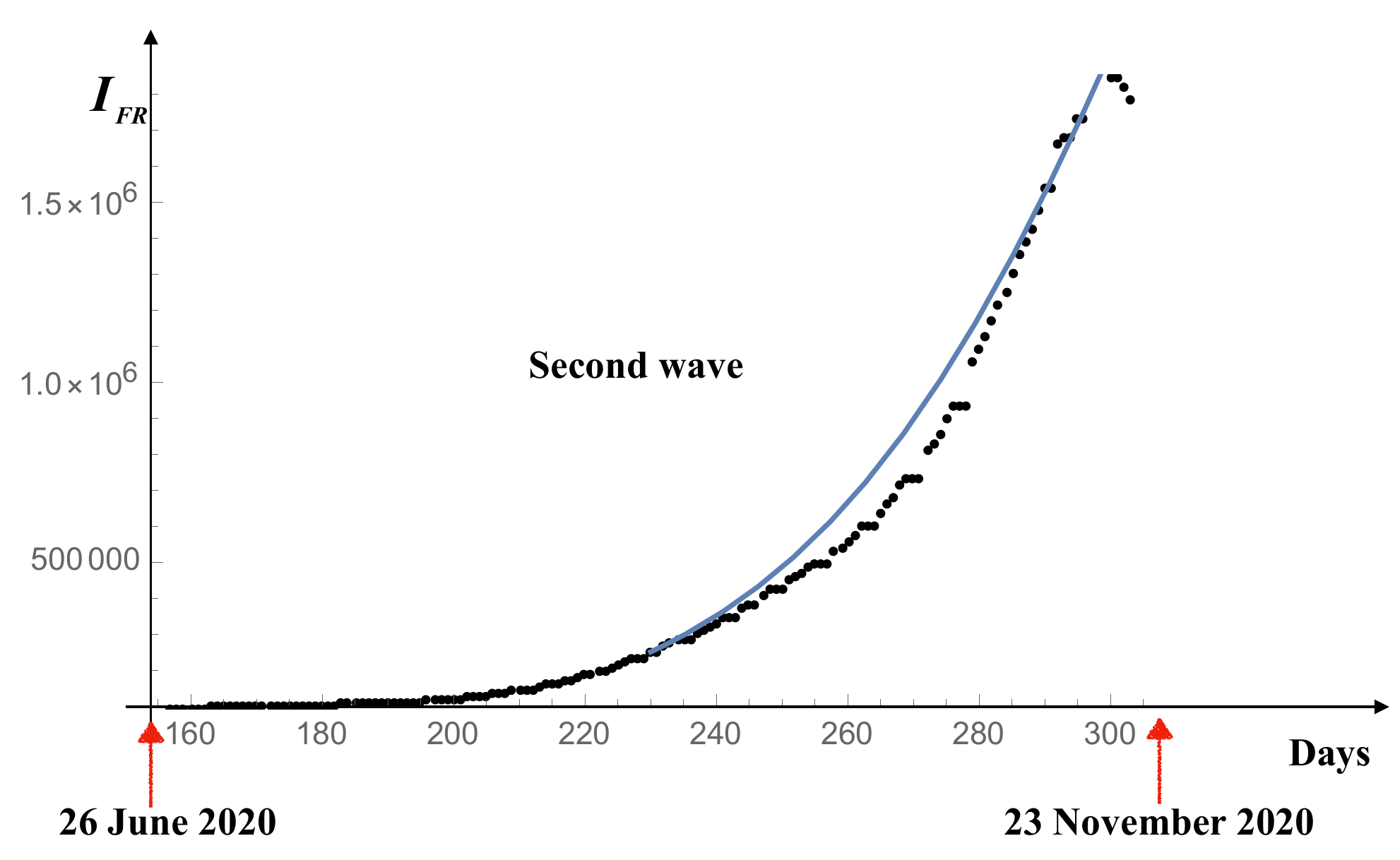}
      \caption{
\textit{Comparison between the theoretical predictions (blue line) and real data (black dots) for France during the second wave of Coronavirus. The values of the parameters are; ${\tilde\alpha}=0.035\ {\rm day}^{-1}$, $\beta=0.5\ {\rm day}^{-2}$, $K=5000000$, and $t_L=278\ {\rm days}$, respectively.}
      }
      \label{Francesw}
    \end{minipage}
  \hfill
\end{figure*}

\noindent Now, let us deal with the stochastic processes. Fig.~\ref{FRfwStoc}. and Fig.~\ref{FRswStoc}. illustrate the comparison between the theoretical predictions (blue lines) and real data (black dots) for France concerning the first and the second waves of SARS-CoV-2, respectively. The intensity of the noise has been estimated by using Eq.~(\ref{comp3}). As we can see, the predictions of our model are in a fairly good agreement with real data. 
\begin{figure*}[htb]
  \hfill
  \begin{minipage}[t]{.48\textwidth}
    \centering
    \includegraphics[width=5cm,height=5cm]{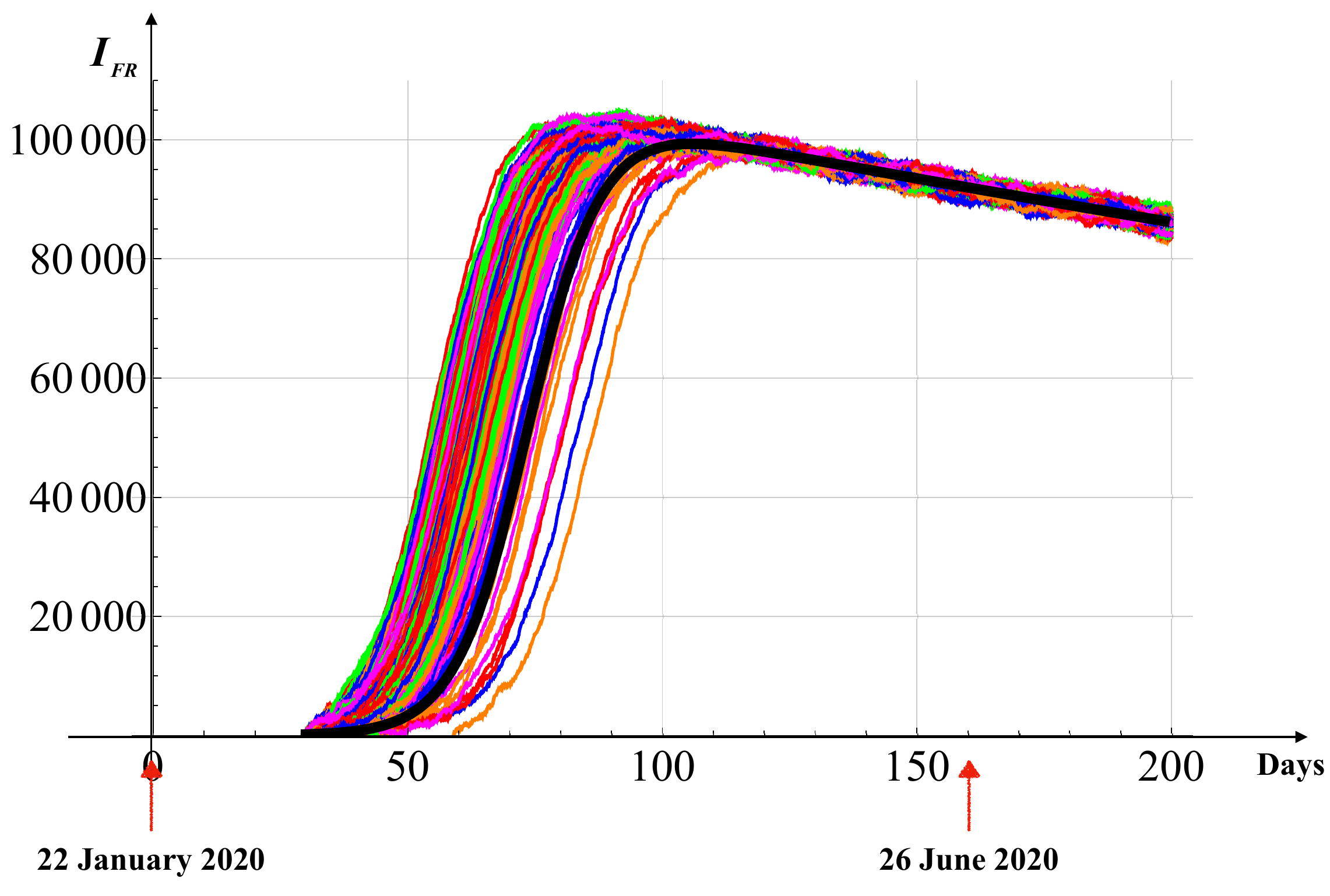}
    \caption{\textit{{\bf Solutions (200 realisations) of the stochastic differential equation for France - First wave}. The values of the parameters are those reported in Fig.~\ref{Francefw}. ${\tilde\eta}=0.002$, which has been estimated according to Eq.~(\ref{comp3}). The black curve is the solution of the deterministic equation.}}
    \label{FRfwStoc}
  \end{minipage}
  \hfill
    \begin{minipage}[t]{.48\textwidth}
      \centering
      \includegraphics[width=5cm,height=5cm]{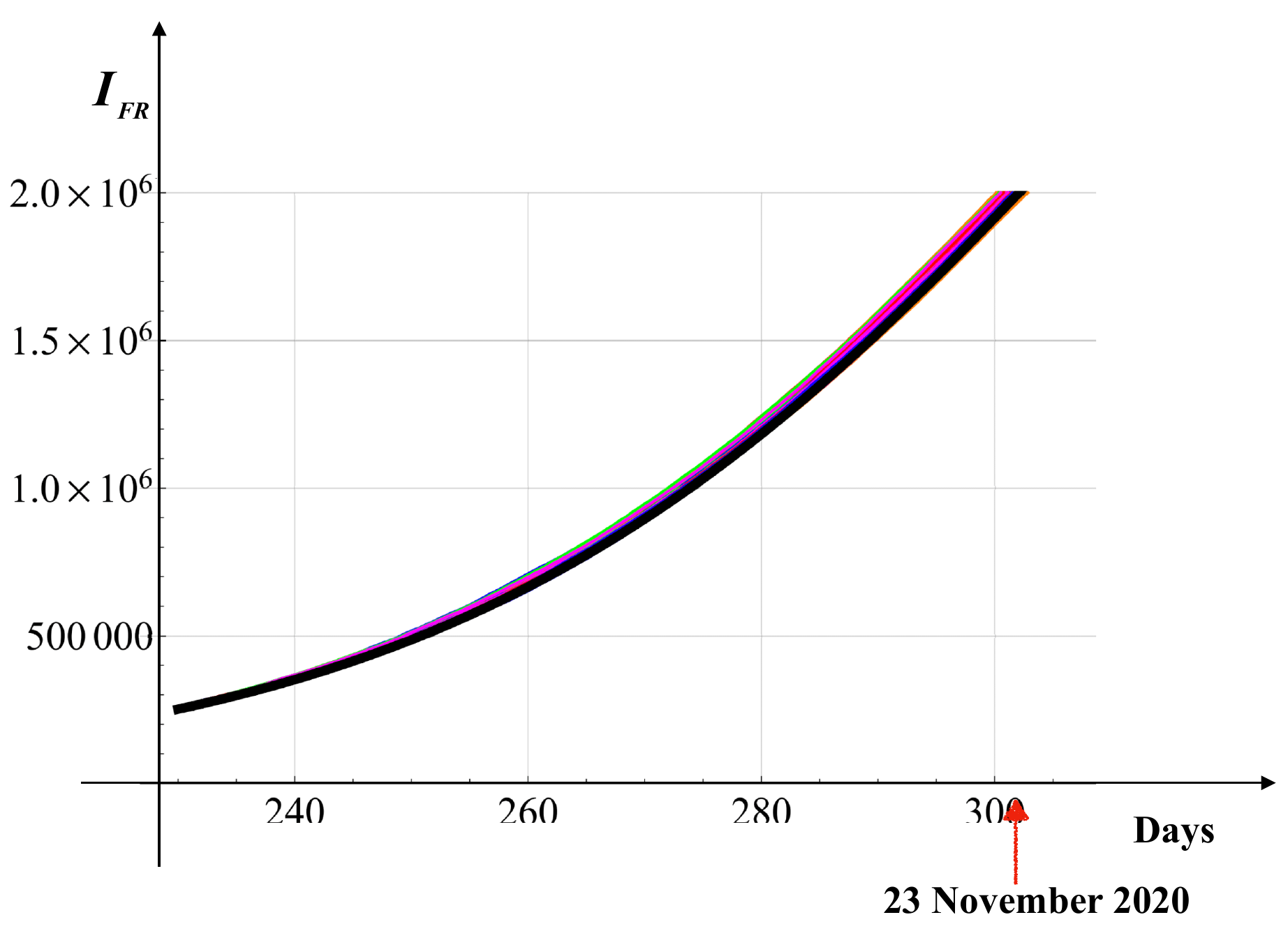}
      \caption{\textit{{\bf Solutions (200 realisations) of the stochastic differential equation for France - Second wave}. The values of the parameters are those reported in Fig.~\ref{Francesw}. ${\tilde\eta}=0.001$, which has been estimated according to Eq.~(\ref{comp3}).}}
      \label{FRswStoc}
    \end{minipage}
  \hfill
\end{figure*}

\noindent As for US, we have analised two possible hypotheses:

\noindent {\bf 1)} USA is still subject to the first wave of Coronavirus infection;

\noindent {\bf 2)} USA is in the second or third wave of SARS-CoV-2 infection.

\noindent Figs.~\ref{USfwStoc}. and \ref{US3w1Stoc}. illustrate these two scenarios. More specifically, Fig.~\ref{USfwStoc}. refers to scenario~{\bf 1)}. In order to obtain the best possible agreement with real data, we have increased, arbitrarily, the intensity of the noise up to a value that is $45$ times higher than that estimated by Eq.~(\ref {comp3}) (we set ${\tilde\eta}=0.009$). However, we did not reach this objective is a satisfactory way. 

\noindent Fig.~\ref{US3w1Stoc}. refers to scenario~{\bf 2)}. In this case the intensity of the noise corresponds to the value estimated by Eq.~(\ref{comp3}) (we have ${\tilde\eta}=0.0002$). Here, we get a fairly good agreement with real data. Finally, our results lead to two possible interpretations:

\noindent {\bf a)} USA is experiencing a second (or even third) wave of Coronavirus infection;

\noindent {\bf b)} USA is still subject to the first wave of infection. However, given the heterogeneity and vastness of the territory, the variables of the system (i.e. $S$, $I$, $R$, and $D$) must depend on time as well as on space. In this case, the correct equations governing the evolution of the compartments must be of Stochastic Partial Differential Equations (S.P.D.E.s)-type.

\noindent In our opinion, scenario {\bf b)} is the correct interpretation.

\begin{figure*}[htb]
  \hfill
  \begin{minipage}[t]{.48\textwidth}
    \centering
    \includegraphics[width=5cm,height=5cm]{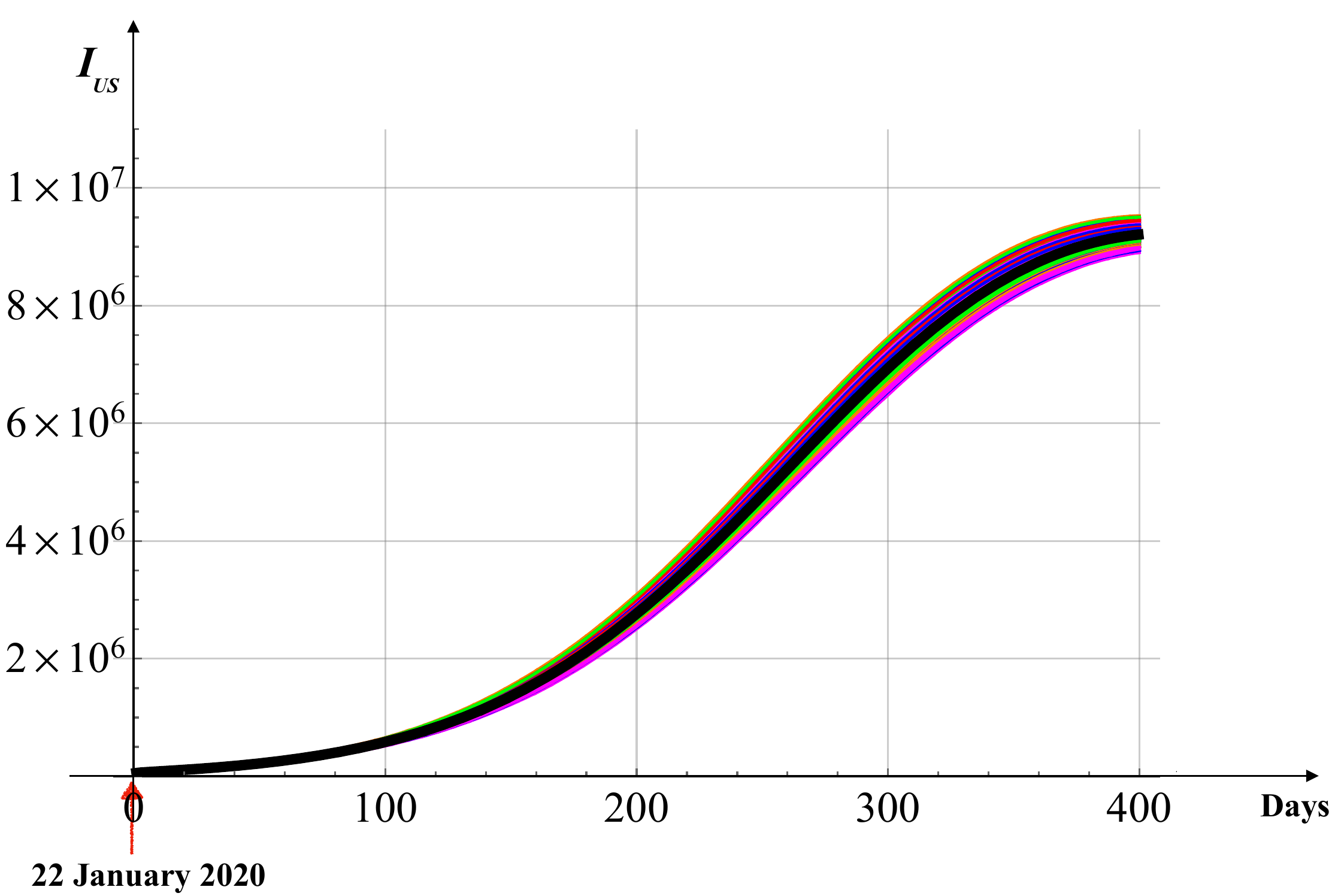}
    \caption{\textit{{\bf Solutions (200 realisations) of the stochastic differential equation for USA - First-Wave Model}. The values of the parameters are those reported in Fig.~\ref{USAfw}. The intensity of the noise, has been set arbitrarily at the value ${\tilde\eta}=0.009$, i.e. $45$ times greater than the value estimated by Eq.~(\ref{comp3}). }}
    \label{USfwStoc}
  \end{minipage}
  \hfill
    \begin{minipage}[t]{.48\textwidth}
      \centering
      \includegraphics[width=5cm,height=5cm]{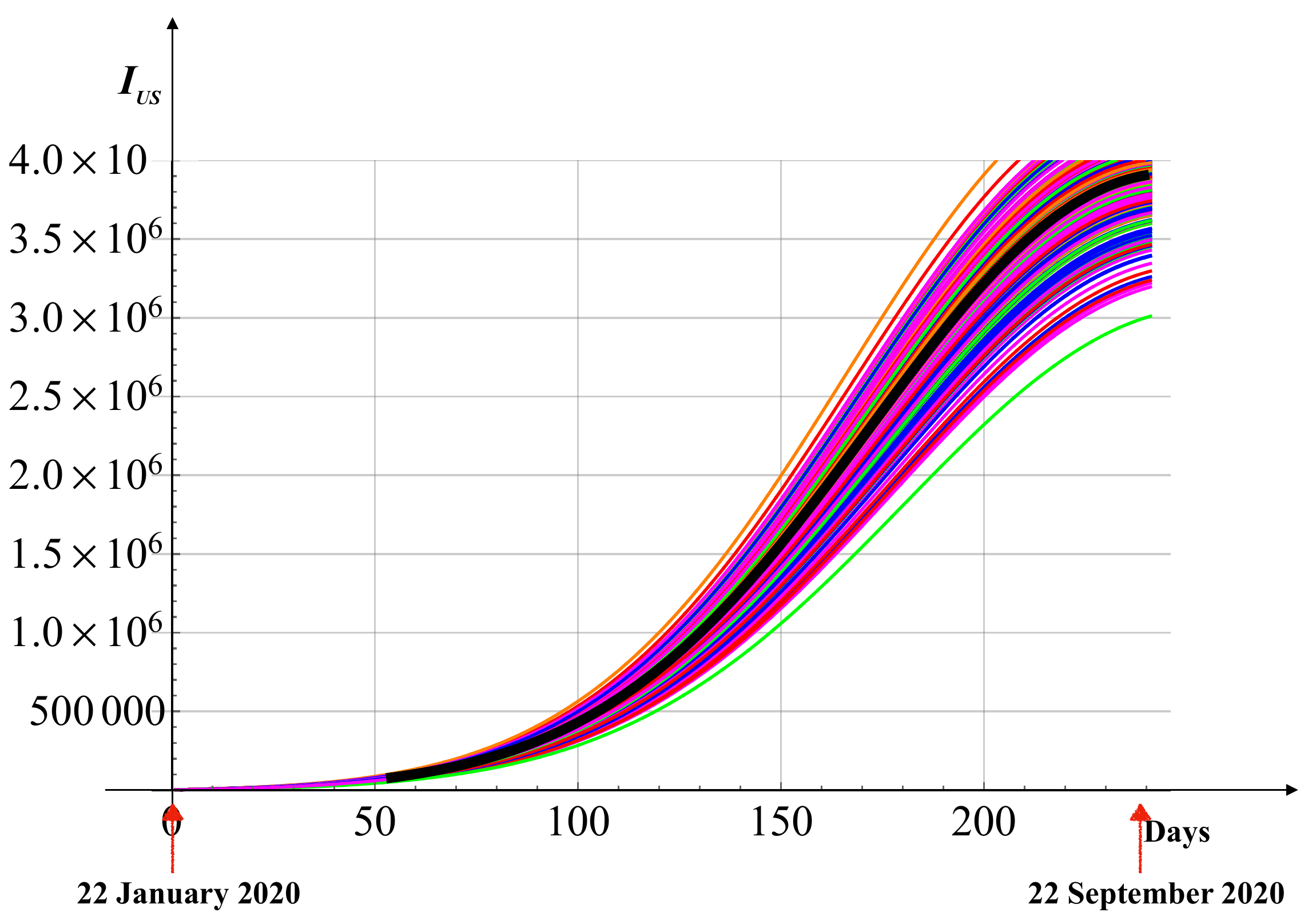}
      \caption{\textit{{\bf Solutions (200 realisations) of the stochastic differential equation for USA - Three-Waves Model}. The values of the parameters are those reported in Fig.~\ref{USA3w}. The intensity of the noise is ${\tilde\eta}=0.0002$, which has been obtained by using Eq.~(\ref{comp3}). The black curve is the solution of the deterministic equation.}}
      \label{US3w1Stoc}
    \end{minipage}
  \hfill
\end{figure*}

\section{\bf Conclusions}\label{C}

\noindent Waiting for the mass production of the vaccine, we studied two scenarios by means of two mathematical models: the $(SIS)_L$-model and the $(SISI_hr_hd_h)_L$-model. 

\noindent The first scenario, described by the $(SIS)_L$-model, refers to the situation in which the role of Hospitals is not taken into account. The second model (the $(SISI_hr_hd_h)_L$-model) integrates the contribution provided by the Hospitals and Health Care Institutes. The dynamics are governed by stochastic differential equations, which are suitable for describing realistic situations. We compute the relevant correlation functions as well as the probability density functions. 

\noindent In the ${SIS}_L$ model, we determined the minimum value of infectious individuals that must be reached after the lockdown measures so that, once these measures are removed, during the time interval $\Delta{\hat t}$}, the number of infectious still stay below a set threshold.

\noindent The $(SISI_hr_hd_h)_L$-model shows the crucial role played by the Hospitals. More specifically, we showed that the Health Care Institutions can influence the outcome of the outbreak, limiting, and even dampening, the spread of the Coronavirus. By way of example, we dealt with the cases of the United States and France. The comparison between the theoretical predictions with real data confirmed the validity of our models. In particular, for US, we have examined two possible scenarios: USA is still subject to the first wave of infection by Coronavirus and USA is in the second (or third) wave of SARS-CoV-2 infection. We concluded that, for a so vast and heterogeneous Country, spatial dependance of the variables can no longer be ignored and the correct equations governing the evolution of the different \textit{compartments}, $S$, $I$, $R$, and $D$, should be of S.P.D.E.s-type. 
\vskip0.5cm
\section{Acknowledgments}\label{ack}

\noindent GS is grateful to Prof. M. Malek Mansour, of the Universit{\'e} Libre de Bruxelles (ULB), for his useful suggestions.
\vskip0.5truecm
\section{\bf APPENDIX 1: A Rough Estimate of the Intensity of the Noise}\label{A1}

\noindent The intensity of the noise for the entire process may be estimated by adopting the concept of \textit{Grand Ensemble} in statistical mechanics. More specifically, we may establish the following correspondences: \textit{Quantity of SARS-CoV-2} corresponds to the \textit{Energy of the system}, the \textit{Quantity of SARS-CoV-2 that can be absorbed or released due to a change of the Infectious individuals} models the \textit{Chemical Potential} ($\mu^{ch.}$), and the \textit{Surface of a Country} corresponds to the \textit{Volume of the system}, respectively. \textit{Temperature of a system} (T) is mimicked by the \textit{surrounding environment}, which is more or less favourable to the spread of Coronavirus depending on whether particular prevention measures are adopted (e.g. use of masks, maintaining a safe distance, tracking of the infection, adoption of lockdown, quarantine and isolation measures, etc.). Hence, in this scenario:

\noindent
{\bf i)} \textit{Temperature equilibrium} ("$T_{Country}=T_{Env.}$") stands for \textit{Environment equilibrium};

\noindent 
{\bf ii)} Identical chemical potentials ($\mu^{ch.}_{Country}=\mu^{ch.}_{Env.}$) stands for equal variation of the quantity of Coronavirus with respect to a change of the Infectious people;

\noindent 
{\bf iii)} Of course the \textit{Surface of a Country} (corresponding to the \textit{Volume of the system}) is kept constant;

\noindent
{\bf iv)} The total number of Infectious individuals (corresponding to the "\textit{number of particles}") and the quantity of the SARS-CoV-2 in a Country (corresponding to the "\textit{Energy of the system}") are not fixed.

\noindent Under the above conditions, we may consider that the system is governed by the \textit{Grand Ensemble} statistical mechanics. So, according to this statistics, the relative fluctuations of the  number of the infectious people behaves as $K_{Country}^{-1/2}$ \cite{reif}, with $K_{Country}$ denoting the capacity  of the Country's population. It is easily checked that this implies that the intensity of the noise in Eq.~(\ref{S1}) is of the order of 
\begin{equation}\label{A11}
\eta\sim t_L K_{Country}^{-1/2}
\end{equation}
\noindent with $t_L$ denoting the time when the lockdown measures have been applied. We may object that this reasoning is based on \textit{equilibrium conditions}. However, we would like to point out that the aim of this Section is only to provide a (rough) estimate of the order of magnitude of the noise.

\vskip0.5cm
\section{\bf APPENDIX 2: Exact Solution of Richards' differential Equation with the Lockdown Term}\label{A2}

\noindent The Richards curve is an extension of the logistic function, allowing for more flexible $S$-shaped curves. This curve satisfies the O.D.E. \cite{Richards}
\begin{equation}\label{A1}
\frac{dI}{dt}=\alpha I\left(1-\left(\frac{I}{K(t)}\right)^\nu\right)
\end{equation}
\noindent with $\nu >0$ (with $\nu >0$) denoting a parameter affecting the point where the asymptote maximum growth occurs. In presence of the therapy-induced death rate, the Richards' equation takes the form
\begin{equation}\label{A2}
\frac{dI}{dt}=\alpha I\left(1-\left(\frac{I}{K(t)}\right)^\nu\right)-c(t)I
\end{equation}
\noindent The goal is to determine the analytic solution of Eq.~(\ref{A2})\footnote{We report the procedure to find the exact solution of Eq.~(\ref{A2}) since it is not easy to find it in literature.}. Let us first rewrite Eq.~(\ref{A2}) as
\begin{equation}\label{A2a}
\frac{1}{I}\frac{dI}{dt}=\alpha\left(1-\left(\frac{I}{K(t)}\right)^\nu\right)-c(t)
\end{equation}
\noindent Successively, we introduce the new functions $n(t)$ and $C(t)$ defined as
\begin{equation}\label{A3}
n(t)\equiv\ln(I(t))+C(t)\qquad {\rm where}\qquad C(t)\equiv\int_{t_0}^t c(s)ds
\end{equation}
\noindent with $t_0$ denoting the initial time. We get
\begin{align}\label{A4}
&\frac{dn}{dt}=\frac{1}{I}\frac{dI}{dt}+\frac{dC}{dt}=\frac{1}{I}\frac{dI}{dt}+c(t)\qquad{\rm and}\\
&\frac{d}{dt}(n(t)-\alpha t)=-\alpha\exp(\nu (n(t)-\alpha t))\frac{\exp(-\nu (C(t)-\alpha t))}{K(t)^\nu}\nonumber
\end{align}
\noindent By setting
\begin{equation}\label{A5}
m(t) \equiv n(t)-\alpha t
\end{equation}
\noindent we obtain
\begin{equation}\label{A6}
\int_{m_0}^m\frac{dm}{\exp(\nu m)}= -\alpha\int_{t_0}^t\frac{\exp(-\nu(C(s)-\alpha s)}{K(s)^\nu}ds
\end{equation}
\noindent which, after integration, reads
\begin{equation}\label{A7}
\exp(m(t))=\left(c_0+\nu\alpha\int_{t_0}^t\frac{\exp(-\nu(C(s)-\alpha s)}{K(s)^\nu}ds\right)^{-1/\nu}
\end{equation}
\noindent where $c_0$ is a constant independent of time. Finally, we get
\begin{equation}\label{A8}
I(t)= \exp(\alpha t-C(t))\left(c_0+\nu\alpha\int_{t_0}^t\frac{\exp(-\nu(C(s)-\alpha s)}{K(s)^\nu}ds\right)^{-1/\nu}
\end{equation}
\noindent Solution~(\ref{A8}) may be cast in the following, more convenient, form
\begin{equation}\label{A9}
I=I_0\frac{\exp(\alpha (t-t_0)-C(t))}{\!\left(1+\nu\alpha I_0^\nu\exp(-\alpha\nu t_0)\int_{t_0}^t(\exp(-\nu(C(s)-\alpha s))/K(s)^\nu ds\right)^{1/\nu}}
\end{equation}
\noindent with $I_0$ denoting the number of population $I$ at time $t_0$ ($I_0\equiv I(t_0)$).
\vskip0.5truecm
\section{\bf APPENDIX 3: Tables of Infectious individuals by SARS-CoV-2 - USA and France}\label{A2}

\begin{table}[t]
\centering
\footnotesize
\begin{tabular}{lrrrr}
  \toprule
  Date & USA& France  \\
  \midrule
22-Jan	&	1	&	0	\\
23-Jan	&	1	&	0	\\
24-Jan	&	2	&	2	\\
25-Jan	&	2	&	3	\\
26-Jan	&	5	&	3	\\
27-Jan	&	5	&	3	\\
28-Jan	&	5	&	4	\\
29-Jan	&	6	&	5	\\
30-Jan	&	6	&	5	\\
31-Jan	&	8	&	5	\\
1-Feb	&	8	&	6	\\
2-Feb	&	8	&	6	\\
3-Feb	&	11	&	6	\\
4-Feb	&	11	&	6	\\
5-Feb	&	11	&	6	\\
6-Feb	&	12	&	6	\\
7-Feb	&	12	&	6	\\
8-Feb	&	12	&	11	\\
9-Feb	&	9	&	11	\\
10-Feb	&	9	&	11	\\
11-Feb	&	10	&	11	\\
12-Feb	&	10	&	9	\\
13-Feb	&	11	&	9	\\
14-Feb	&	11	&	9	\\
15-Feb	&	11	&	7	\\
16-Feb	&	11	&	7	\\
17-Feb	&	11	&	7	\\
18-Feb	&	11	&	7	\\
19-Feb	&	11	&	7	\\
20-Feb	&	11	&	7	\\
21-Feb	&	11	&	7	\\
22-Feb	&	11	&	7	\\
23-Feb	&	11	&	7	\\
24-Feb	&	11	&	7	\\
25-Feb	&	10	&	2	\\
26-Feb	&	10	&	5	\\
27-Feb	&	11	&	25	\\
28-Feb	&	10	&	44	\\
29-Feb	&	17	&	86	\\
1-Mar	&	24	&	116	\\
2-Mar	&	42	&	176	\\
3-Mar	&	60	&	188	\\
4-Mar	&	89	&	272	\\
5-Mar	&	165	&	362	\\
6-Mar	&	216	&	635	\\
7-Mar	&	379	&	936	\\
8-Mar	&	491	&	1105	\\
9-Mar	&	565	&	1188	\\
10-Mar	&	746	&	1749	\\
11-Mar	&	1106	&	2233	\\
12-Mar	&	1531	&	2233	\\
13-Mar	&	2156	&	3590	\\
14-Mar	&	2908	&	4393	\\
15-Mar	&	3130	&	4429	\\
 \bottomrule
\end{tabular}
\caption{\footnotesize\textit{Number of infectious individuals by SARS-CoV2 in USA and France}.}
\label{A}
\end{table}

\begin{table}[t]
\centering
\footnotesize
\begin{tabular}{lrrrr}
  \toprule
  Date & USA& France  \\
  \midrule
16-Mar	&	4565	&	6522	\\
17-Mar	&	6363	&	7554	\\
18-Mar	&	8873	&	8963	\\
19-Mar	&	13278	&	10714	\\
20-Mar	&	19518	&	12295	\\
21-Mar	&	25385	&	13888	\\
22-Mar	&	34103	&	13361	\\
23-Mar	&	45146	&	17055	\\
24-Mar	&	55347	&	18270	\\
25-Mar	&	67083	&	20360	\\
26-Mar	&	84191	&	22898	\\
27-Mar	&	102165	&	25698	\\
28-Mar	&	120937	&	30064	\\
29-Mar	&	137630	&	30871	\\
30-Mar	&	156037	&	34176	\\
31-Mar	&	179881	&	39782	\\
1-Apr	&	203134	&	42282	\\
2-Apr	&	231492	&	41983	\\
3-Apr	&	261362	&	44547	\\
4-Apr	&	287888	&	46354	\\
5-Apr	&	311723	&	46970	\\
6-Apr	&	337987	&	48989	\\
7-Apr	&	364651	&	49297	\\
8-Apr	&	392895	&	50718	\\
9-Apr	&	423395	&	51725	\\
10-Apr	&	451289	&	53328	\\
11-Apr	&	475982	&	54349	\\
12-Apr	&	499386	&	79831	\\
13-Apr	&	512755	&	82407	\\
14-Apr	&	534593	&	85829	\\
15-Apr	&	557816	&	85461	\\
16-Apr	&	584439	&	95341	\\
17-Apr	&	611175	&	94981	\\
18-Apr	&	630837	&	92778	\\
19-Apr	&	649097	&	96713	\\
20-Apr	&	673345	&	97717	\\
21-Apr	&	693988	&	98189	\\
22-Apr	&	718702	&	93981	\\
23-Apr	&	746583	&	94357	\\
24-Apr	&	759298	&	94211	\\
25-Apr	&	787874	&	94433	\\
26-Apr	&	806768	&	94458	\\
27-Apr	&	824054	&	97154	\\
28-Apr	&	841962	&	98504	\\
29-Apr	&	862462	&	94225	\\
30-Apr	&	856306	&	93444	\\
1-May	&	878601	&	92496	\\
2-May	&	893311	&	93092	\\
3-May	&	911895	&	93140	\\
4-May	&	926931	&	92903	\\
5-May	&	946531	&	92308	\\
6-May	&	969157	&	94333	\\
7-May	&	989843	&	93737	\\
8-May	&	1010883	&	94077	\\
9-May	&	1020748	&	94321	\\
10-May	&	1035084	&	94384	\\
  \bottomrule
\end{tabular}
\caption{\footnotesize\textit{Number of infectious individuals by SARS-CoV2 in USA and France}.}
\label{A}
\end{table}

\begin{table}[t]
\centering
\footnotesize
\begin{tabular}{lrrrr}
  \toprule
  Date & USA& France  \\
  \midrule
11-May	&	1036945	&	94066	\\
12-May	&	1060502	&	93457	\\
13-May	&	1066573	&	92321	\\
14-May	&	1088973	&	91847	\\
15-May	&	1108231	&	91536	\\
16-May	&	1113346	&	91536	\\
17-May	&	1127017	&	90255	\\
18-May	&	1136924	&	89966	\\
19-May	&	1150239	&	90230	\\
20-May	&	1166863	&	90093	\\
21-May	&	1187411	&	89757	\\
22-May	&	1157787	&	89735	\\
23-May	&	1166628	&	89694	\\
24-May	&	1180460	&	89604	\\
25-May	&	1185875	&	89290	\\
26-May	&	1198869	&	88317	\\
27-May	&	1209288	&	87737	\\
28-May	&	1222490	&	90390	\\
29-May	&	1239114	&	90285	\\
30-May	&	1252000	&	91592	\\
31-May	&	1242033	&	91731	\\
1-Jun	&	1244711	&	91954	\\
2-Jun	&	1259414	&	90709	\\
3-Jun	&	1263066	&	90337	\\
4-Jun	&	1277666	&	90539	\\
5-Jun	&	1295466	&	90576	\\
6-Jun	&	1307001	&	90822	\\
7-Jun	&	1318704	&	91115	\\
8-Jun	&	1323590	&	91051	\\
9-Jun	&	1334677	&	90924	\\
10-Jun	&	1346078	&	91120	\\
11-Jun	&	1361476	&	91201	\\
12-Jun	&	1378478	&	91474	\\
13-Jun	&	1393707	&	91740	\\
14-Jun	&	1407202	&	92087	\\
15-Jun	&	1411814	&	92024	\\
16-Jun	&	1427516	&	91966	\\
17-Jun	&	1444596	&	92064	\\
18-Jun	&	1464830	&	92283	\\
19-Jun	&	1487551	&	92850	\\
20-Jun	&	1508109	&	93280	\\
21-Jun	&	1529107	&	93497	\\
22-Jun	&	1541333	&	93607	\\
23-Jun	&	1569358	&	93808	\\
24-Jun	&	1594422	&	93622	\\
25-Jun	&	1626842	&	93377	\\
26-Jun	&	1664318	&	94641	\\
27-Jun	&	1696670	&	94641	\\
28-Jun	&	1730945	&	93921	\\
29-Jun	&	1750571	&	95582	\\
30-Jun	&	1780503	&	95818	\\
1-Jul	&	1821928	&	96443	\\
2-Jul	&	1824839	&	96835	\\
3-Jul	&	1867592	&	97141	\\
4-Jul	&	1808946	&	97141	\\
5-Jul	&	1845794	&	97141	\\  \bottomrule
\end{tabular}
\caption{\footnotesize\textit{Number of infectious individuals by SARS-CoV2 in USA and France}.}
\label{A}
\end{table}

\begin{table}[t]
\centering
\footnotesize
\begin{tabular}{lrrrr}
  \toprule
  Date & USA& France  \\
  \midrule
6-Jul	&	1872210	&	98230	\\
7-Jul	&	1919330	&	98356	\\
8-Jul	&	1961331	&	98356	\\
9-Jul	&	2007257	&	99079	\\
10-Jul	&	2060216	&	99495	\\
11-Jul	&	2107164	&	99495	\\
12-Jul	&	2155023	&	99495	\\
13-Jul	&	2187840	&	100886	\\
14-Jul	&	2237159	&	100886	\\
15-Jul	&	2277035	&	101500	\\
16-Jul	&	2338454	&	101800	\\
17-Jul	&	2392640	&	102417	\\
18-Jul	&	2438729	&	102417	\\
19-Jul	&	2490585	&	102417	\\
20-Jul	&	2522582	&	104175	\\
21-Jul	&	2563770	&	104578	\\
22-Jul	&	2605612	&	105346	\\
23-Jul	&	2650612	&	105882	\\
24-Jul	&	2694315	&	106659	\\
25-Jul	&	2740987	&	106661	\\
26-Jul	&	2777024	&	106661	\\
27-Jul	&	2803934	&	108928	\\
28-Jul	&	2838789	&	109410	\\
29-Jul	&	2875126	&	109408	\\
30-Jul	&	2916772	&	110596	\\
31-Jul	&	2959434	&	113165	\\
1-Aug	&	2992425	&	113166	\\
2-Aug	&	3031480	&	113166	\\
3-Aug	&	3030750	&	113166	\\
4-Aug	&	3071148	&	116515	\\
5-Aug	&	3075303	&	115981	\\
6-Aug	&	3112757	&	118410	\\
7-Aug	&	3144584	&	121912	\\
8-Aug	&	3178688	&	121913	\\
9-Aug	&	3210796	&	121939	\\
10-Aug	&	3245109	&	126051	\\
11-Aug	&	3246746	&	126056	\\
12-Aug	&	3263231	&	130104	\\
13-Aug	&	3293106	&	130092	\\
14-Aug	&	3334854	&	135252	\\
15-Aug	&	3358391	&	138562	\\
16-Aug	&	3384449	&	138562	\\
17-Aug	&	3388006	&	141885	\\
18-Aug	&	3399126	&	141886	\\
19-Aug	&	3418273	&	141886	\\
20-Aug	&	3439319	&	141886	\\
21-Aug	&	3468440	&	156471	\\
22-Aug	&	3490679	&	159943	\\
23-Aug	&	3512398	&	164816	\\
24-Aug	&	3525433	&	166529	\\
25-Aug	&	3531255	&	169596	\\
\bottomrule
\end{tabular}
\caption{\footnotesize\textit{Number of infectious individuals by SARS-CoV2 in USA and France}.}
\label{A}
\end{table}

\begin{table}[t]
\centering
\footnotesize
\begin{tabular}{lrrrr}
  \toprule
  Date & USA& France  \\
  \midrule
26-Aug	&	3544757	&	175014	\\
27-Aug	&	3571899	&	180633	\\
28-Aug	&	3600855	&	187851	\\
29-Aug	&	3622993	&	187851	\\
30-Aug	&	3644136	&	198707	\\
31-Aug	&	3647867	&	201556	\\
1-Sep	&	3670637	&	206266	\\
2-Sep	&	3681431	&	212950	\\
3-Sep	&	3688945	&	219847	\\
4-Sep	&	3721623	&	228610	\\
5-Sep	&	3745253	&	228610	\\
6-Sep	&	3762544	&	228611	\\
7-Sep	&	3768258	&	247958	\\
8-Sep	&	3769148	&	254072	\\
9-Sep	&	3773487	&	263245	\\
10-Sep	&	3792628	&	271956	\\
11-Sep	&	3824476	&	281099	\\
12-Sep	&	3847957	&	281464	\\
13-Sep	&	3865128	&	281545	\\
14-Sep	&	3875410	&	304480	\\
15-Sep	&	3892945	&	312082	\\
16-Sep	&	3899974	&	321520	\\
17-Sep	&	3929266	&	331398	\\
18-Sep	&	3960964	&	343464	\\
19-Sep	&	3981720	&	342771	\\
20-Sep	&	4006695	&	342771	\\
21-Sep	&	4032529	&	371216	\\
22-Sep	&	4039967	&	380763	\\
23-Sep	&	4054155	&	380511	\\
24-Sep	&	4059018	&	408785	\\
25-Sep	&	4089001	&	423837	\\
26-Sep	&	4109430	&	423842	\\
27-Sep	&	4130607	&	423861	\\
28-Sep	&	4134248	&	449109	\\
29-Sep	&	4157434	&	460025	\\
30-Sep	&	4170312	&	473413	\\
1-Oct	&	4194775	&	485798	\\
2-Oct	&	4235533	&	497007	\\
3-Oct	&	4259573	&	497057	\\
4-Oct	&	4280939	&	497057	\\
5-Oct	&	4295944	&	531099	\\
6-Oct	&	4321937	&	541910	\\
7-Oct	&	4324142	&	559079	\\
8-Oct	&	4360138	&	576623	\\
9-Oct	&	4397371	&	596601	\\
10-Oct	&	4426857	&	596601	\\
11-Oct	&	4460175	&	596601	\\
12-Oct	&	4468893	&	639359	\\
13-Oct	&	4503100	&	660072	\\
14-Oct	&	4530389	&	680944	\\
15-Oct	&	4572309	&	710808	\\
  \bottomrule
\end{tabular}
\caption{\footnotesize\textit{Number of infectious individuals by SARS-CoV2 in USA and France}.}
\label{A}
\end{table}

\begin{table}[t]
\centering
\footnotesize
\begin{tabular}{lrrrr}
  \toprule
  Date & USA& France  \\
  \midrule
16-Oct	&	4619704	&	735003	\\
17-Oct	&	4652793	&	735003	\\
18-Oct	&	4688096	&	735003	\\
19-Oct	&	4716172	&	809342	\\
20-Oct	&	4753794	&	828832	\\
21-Oct	&	4787492	&	854579	\\
22-Oct	&	4832242	&	895092	\\
23-Oct	&	4889906	&	936487	\\
24-Oct	&	4940522	&	936512	\\
25-Oct	&	4985683	&	936512	\\
26-Oct	&	5014178	&	1058635	\\
27-Oct	&	5061982	&	1091260	\\
28-Oct	&	5109189	&	1124979	\\
29-Oct	&	5162232	&	1171071	\\
30-Oct	&	5235613	&	1218518	\\
31-Oct	&	5289578	&	1252788	\\
1-Nov	&	5346628	&	1298278	\\
2-Nov	&	5385178	&	1351392	\\
3-Nov	&	5478077	&	1387595	\\
4-Nov	&	5541476	&	1426229	\\
5-Nov	&	5628278	&	1480142	\\
6-Nov	&	5723823	&	1537990	\\
7-Nov	&	5810136	&	1536134	\\
8-Nov	&	5889532	&	1660602	\\
9-Nov	&	5962848	&	1680027	\\
10-Nov	&	6066108	&	1681031	\\
11-Nov	&	6172962	&	1733142	\\
12-Nov	&	6278908	&	1733471	\\
13-Nov	&	6411436	&	1785521	\\
14-Nov	&	6523581	&	1815277	\\
15-Nov	&	6629769	&	1841533	\\
16-Nov	&	6719435	&	1849775	\\
17-Nov	&	6829398	&	1892168	\\
18-Nov	&	6940785	&	1917979	\\
19-Nov	&	7066805	&	1936750	\\
20-Nov	&	7213710	&	1956173	\\
21-Nov	&	7318556	&	1972675	\\
22-Nov	&	7463554	&	1985114	\\
23-Nov	&	7524768	&	1988458	\\
  \bottomrule
\end{tabular}
\caption{\footnotesize\textit{Number of infectious individuals by SARS-CoV2 in USA and France}.}
\label{A}
\end{table}

\end{document}